\pdfoutput=1
\documentclass[cernpreprint,american,orcidlogo,atlasdraft=false, texmf, orcidlogo]{atlasdoc}
 
\usepackage[subfigure]{atlaspackage}
\usepackage{atlasbiblatex}
 
\usepackage{atlasphysics}
 
\graphicspath{{logos/}{figures/}}
 
\usepackage{ANA-STDM-2017-37-PAPER-defs}
\renewcommand{\todo}[2][]{}

\AtlasTitle{Measurement of cross-sections for production of a \Z boson in
association with a flavor-inclusive or doubly $b$-tagged large-radius jet in
proton-proton collisions at $\sqrt{s} = 13~\TeV$ with the ATLAS experiment}

\AtlasAbstract{
We present measurements of cross-sections for production of a
leptonically decaying \Z boson in association with a large-radius jet in
13~\TeV{} proton-proton collisions at the LHC, using $36~\text{fb}^{-1}$ of
data from the ATLAS detector.
Integrated and differential
cross-sections are measured at particle-level in both a flavor-inclusive and
a doubly $b$-tagged fiducial phase-space.  The large-radius jet mass and transverse
momentum, its kinematic relationship to the \Z boson, and the angular
separation of $b$-tagged small-radius track-jets within the large-radius jet are
measured.
This measurement constitutes an
important test of perturbative quantum chromodynamics in kinematic and flavor
configurations relevant to several Higgs boson and beyond-Standard-Model
physics analyses. The results highlight issues with modeling of additional
hadronic activity in the flavor-inclusive selection, and a distinction between
flavor-number schemes in the $b$-tagged phase-space.
}

\AtlasRefCode{STDM-2017-37}
 
\PreprintIdNumber{CERN-EP-2021-140}

\AtlasJournal{Phys.~Rev.~D.}
\AtlasJournalRef{Phys.~Rev.~D 108 (2023) 1, 012022}
\AtlasDOI{10.1103/PhysRevD.108.012022}

\hypersetup{pdftitle={ATLAS document},pdfauthor={The ATLAS Collaboration}}
 
\begin{document}
 
\maketitle

\section{Introduction}
\label{sec:intro}
 
Since the proposal of a $\text{Higgs} \to \bbbar$ discovery channel based on
the structure of a high-momentum jet~\cite{Butterworth:2008iy},
``boosted jet'' methods
have been a major feature of Higgs and other experimental analyses at the
LHC~\cite{Aaboud:2018knk,Aad:2019uoz,Sirunyan:2018qca,Sirunyan:2019quj,Sirunyan:2020hwv,Sirunyan:2020hwz}. In
these, a large-radius jet (\ljet) is reconstructed, and then decomposed into
smaller-radius (small-$R$) subjets whose structure enables identification of
resonances whose decay products have been collimated by their parent's large
momentum.  Such methods are also of high interest in searches for new physics,
both because a high-mass new particle decaying into
resonances naturally generates high-momentum merged jets, and because the
high-momentum regime is particularly sensitive to modifications to Standard
Model (SM) dynamics by new physics.
 
A good understanding of hard, collinear parton splittings in quantum
chromodynamics (QCD) is key to measurements using \ljet methods, as these
constitute the dominant background to boosted resonance signatures.  Monte Carlo
(MC) event-simulation methods include approximations such as factorization of
partonic splittings in parton-shower algorithms, and slicing of emission
phase-space between matrix-element and parton-shower sources. The extreme
dynamics of boosted topologies include aspects of both collinear and high-\pt
physics, presenting a test for such factorizations.
 
The role of heavy-quark masses
is also important: contradictory inclusions of mass effects in calculations for
parton-shower MC generators and parton distribution functions (PDFs) mean that
there is as yet no unambiguously more-correct simulation strategy for
heavy-flavor production~\cite{Napoletano:2019tla}.  Additionally, the modeling of
gluon-splitting into heavy quarks may benefit from renormalization-scale choices
different from those developed for the more common cases of gluon emission from
light quarks and gluons.
 
Due to these factors, empirical comparisons between MC models and collider
measurements of boosted $\bbbar$ production are key to understanding and
improving the validity of parton-shower MC simulations in this event topology,
and closely related ones such as high-\pt associated Higgs-boson production.  In
addition, this phase-space is also relatively novel for light-jet production,
whose modeling can be similarly informed by equivalent measurements without the
$b$-tagging requirement.
 
This analysis follows the $VH(\bbbar)$ event-selection strategy of
requiring boosted-jet production 
in association with a leptonically decaying vector
boson,
which
is effective at reducing QCD backgrounds and provides an additional,
experimentally clean proxy for each event's characteristic momentum-transfer
scale~\cite{Butterworth:2008iy}.
Previous measurements of the \Z{}+$b(\bbar)$ process, in a resolved rather than
boosted event topology, were made
using $\sqrt{s} = 7~\TeV{}$ proton--proton collisions by ATLAS, CMS and LHCb~\cite{STDM-2011-22,STDM-2012-15,
CMS-EWK-11-015,CMS-SMP-13-004,Aaij:2014gta},
at 8~\TeV{} by CMS~\cite{CMS-SMP-14-010}, and at 13~\TeV{}
by ATLAS and CMS~\cite{ATLAS:2020gfi,CMS-SMP-19-004}.
 
Additionally, studies of correlations between $B$ and $\bar{B}$ mesons have been
performed by CMS at 7~\TeV~\cite{CMS-BPH-10-010}, and various
properties of gluon splitting
at 13~\TeV{} have been measured by ATLAS~\cite{STDM-2017-17}. In these papers the differential
cross-section as a function of \drbb, the angular separation between the
$b$-jets or $b$-hadrons, was measured and found to be mismodeled by MC generators in
the small-\drbb region typical of gluon splitting into \bbbar. Given the
importance of \drbb modeling to techniques for reconstruction of boosted Higgs-bosons, this variable is also a
target of this analysis.  The higher 13~\TeV{} center-of-mass energy produces a larger
boosted-event population than in the 7~\TeV{} studies, and in contrast to the recent
ATLAS \Z{}+\bbbar cross-section measurement at 13~\TeV, use is also made of
small-radius charged-particle jets that allow precise measurement of the small
angular separations.
 
In this paper, we present measurements of cross-sections for the production of a
leptonically-decaying \Z-boson
in association with a \ljet, using data taken by the ATLAS detector at the
LHC~\cite{PERF-2007-01,Evans:2008zzb} in 2015--2016. The measured cross-sections
are differential in kinematic variables of the \ljet. An additional phase-space
is defined by the requirement that the \ljet be doubly $b$-tagged: total and
differential cross-sections, including as a function of the angular separation of the $b$-tagged
subjets, are also measured in this phase-space. These measurements provide an
important test of perturbative QCD in the boosted regime, including contributions
both where a high-energy gluon splits to give a \bbbar pair carrying a significant
fraction of the jet momentum, and where secondary processes in which two $b$-tagged
momentum flows exist within the jet but are not its dominant kinematic components.
 
The relevant perturbative-QCD issues are summarized in Section~\ref{sec:theory},
the ATLAS detector is described in Section~\ref{sec:detector}, and the MC event
samples used in the analysis and for comparison with the resulting measurements
are discussed in Section~\ref{sec:samples}. 
The physics-object definitions at the reconstruction and particle levels are
described in Section~\ref{sec:objectdefn}, followed by the event selection and
observable definitions in Section~\ref{sec:eventseln}. The correction for
detector biases is treated in Section~\ref{sec:unfolding}, and the sources and estimation of systematic
uncertainties are described in Section~\ref{sec:systematics}. Finally, the
detector-corrected observables are compared with current MC predictions in
Section~\ref{sec:results}.

\section{Theory context}
\label{sec:theory}
 
Heavy-flavor partons in the initial state of a hard-scattering process are
understood to arise mainly from perturbative gluon splittings into $\bbbar$ and
$\ccbar$ quark--antiquark pairs, formalized as DGLAP QCD
evolution~\cite{gribov1972deep,altarelli1977asymptotic,dokshitzer1977calculation}. But
there is an ambiguity in this evolution in the usual factorization picture of
perturbative QCD calculations, as to whether the emergence of heavy flavor is to be
isolated into the partonic cross-section $\hat{\sigma}$, or is also permitted in
the evolution of the PDFs
which encode the initial-state proton structure. At present, this separation is
strongly tied to the treatment of the heavy quark as having a finite or zero
mass~\cite{Napoletano:2019tla}.
 
The picture without heavy-quark production in the PDF evolution -- here, the
absence of $b$-quarks -- is termed the four-flavor number scheme (4FNS). In
this, the $b$-quark density in the PDF is set to zero, and so the perturbative
generation of initial-state $b$-quarks comes from explicit gluon splitting into a
\bbbar pair in the partonic matrix-element~\cite{Napoletano:2018euk}, usually
including $b$-quark mass effects. A consequence is that in the four-flavor scheme
there are always at least two participating $b$-quarks, although they may fall
outside the experimental acceptance.
By contrast, in the five-flavor number scheme (5FNS) the PDF evolution can
generate initial-state $b$-quarks -- again through gluon splitting, but now
internalized in the functional form of the $b$-quark PDF. This allows matrix-element
amplitudes where it is possible for only one $b$-quark to
participate in the hard-scatter~\cite{Napoletano:2018euk}. While the 5FNS initially
seems the more complete treatment, its treatment of the initial-state $b$-quark
is purely longitudinal (whereas gluon splitting in the matrix element generates
transverse momentum), and to avoid non-cancellation of higher-order soft
divergences the initial-state $b$-quark is currently treated as massless in
standard PDF approaches~\cite{Krauss:2017wmx}.
 
In a hypothetical all-orders calculation the two schemes would give the same
results, but for a truncated perturbation expansion they generally give
different predictions. There are arguments in favour of both approaches: the
4FNS allows for transverse momentum exchange through the initial-state heavy
quarks and hence might be expected to describe event kinematics better, while
the 5FNS is able to make use of higher-order DGLAP resummation calculations in
the PDF evolution, which are not present in matrix elements matched to parton
showers. The importance of mass effects is expected to become less important for
process scales $Q \gg m_b$, suggesting that higher-accuracy predictions may be
obtained using the 5FNS in boosted event
configurations~\cite{Forte:2015hba,Napoletano:2019tla,Bagnaschi:2018dnh,Ridolfi:2019bch}. Recent
computations such as the computation of $\Z+b$ production at
$\mathcal{O}(\alphas^3)$~\cite{Gauld:2020deh}, and the NLO ``fusing'' scheme
developed by \sherpa~\cite{Krauss:2016orf} combine desirable aspects of both
schemes, and should also be dominated by the 5FNS in boosted phase-space. It is
therefore important to compare experimental measurements of $b$-quark production
with predictions using both of these schemes, to empirically test the accuracy
and predictivity of the available theoretical approaches.
 
Theoretical uncertainties also arise in the production of $b$-quarks in the partonic
final state. The usual parton shower formulation for parton or dipole splitting
is derived in the collinear-emission limit, using the \pt of the splitting as
the characteristic (renormalization) scale, but this heuristic scale choice is
only well-motivated for gluon-emission splitting functions~\cite{Ellis:1986bv}.
The scale choice for gluon splitting, especially into heavy quarks, is hence an
extrapolation requiring empirical testing. This is of great importance since
uncertainties in heavy-flavor production by gluon splitting are a leading
systematic limitation on the sensitivity of Higgs boson decays into \bbbar in the
\ttH, \VH (where $V$ is either a \W or \Z boson) and gluon-fusion
channels~\cite{HIGG-2020-23,HIGG-2018-04,CMS:2020zge} --- particularly in
boosted-Higgs configurations where the two $b$-quarks are relatively collinear,
similar to the gluon-splitting kinematics~\cite{HIGG-2018-52}.

\section{ATLAS detector}
\label{sec:detector}
 
The ATLAS detector~\cite{PERF-2007-01} is a multipurpose particle detector with
a forward/backward-symmetric cylindrical geometry. The detector has a nearly
$4\pi$ coverage in solid angle
\footnote{ATLAS uses a right-handed coordinate system with its origin at the
nominal interaction point (IP) in the center of the detector and the $z$-axis
along the beam pipe. The $x$-axis points from the IP to the center of the LHC
ring, and the $y$-axis points upward. Cylindrical coordinates $(r, \phi)$ are
used in the transverse plane, $\phi$ being the azimuthal angle around the beam
pipe. The pseudorapidity is defined in terms of the polar angle $\theta$ as
$\eta = - \ln \tan(\theta/2)$. Angular distance is measured in units of
\(\Delta R \equiv \sqrt{(\Delta\eta)^{2} + (\Delta\phi)^{2}}\).} 
and consists of an inner tracking detector, electromagnetic and hadronic
calorimeters, and a muon spectrometer.
 
The inner-detector system (ID) is immersed in a \SI{2}{\tesla} axial magnetic field
and provides charged-particle tracking in the range \(|\eta| < 2.5\).
The high-granularity silicon pixel detector covers the vertex region and typically provides four measurements per track,
the first hit normally being in the insertable B-layer (IBL) installed before Run~2~\cite{ATLAS-TDR-19,PIX-2018-001}.
It is followed by the silicon microstrip tracker (SCT), which usually provides eight measurements per track.
These silicon detectors are complemented by the transition radiation tracker (TRT),
which enables radially extended track reconstruction up to \(|\eta| = 2.0\).
The TRT also provides electron-identification information
based on the fraction of hits (typically 30 in total) above a higher energy-deposit threshold corresponding to transition radiation.
 
The calorimeter system covers the pseudorapidity range \(|\eta| < 4.9\).
Within the region \(|\eta|< 3.2\), electromagnetic calorimetry is provided by barrel and
endcap high-granularity lead/liquid-argon (LAr) calorimeters,
with an additional thin LAr presampler covering \(|\eta| < 1.8\)
to correct for energy loss in material upstream of the calorimeters.
Hadron calorimetry is provided by the steel/scintillator-tile calorimeter,
segmented into three barrel structures within \(|\eta| < 1.7\), and two copper/LAr hadron endcap calorimeters.
The solid angle coverage is completed with forward copper/LAr and tungsten/LAr calorimeter modules
optimised for electromagnetic and hadronic energy measurements respectively.
 
The muon spectrometer (MS) comprises separate trigger and
high-precision tracking chambers measuring the deflection of muons in a magnetic field generated by the superconducting air-core toroidal magnets.
The field integral of the toroids ranges between \num{2.0} and \SI{6.0}{\tesla\metre}
across most of the detector.
Three layers of precision chambers, each consisting of layers of monitored drift tubes, covers the region \(|\eta| < 2.7\),
complemented by cathode-strip chambers in the forward region, where the background is highest.
The muon trigger system covers the range \(|\eta| < 2.4\) with resistive-plate chambers in the barrel, and thin-gap chambers in the endcap regions.
 
A two-level trigger system is used to select events for further
analysis~\cite{TRIG-2016-01,Aaboud:2016leb}. The first-level trigger is implemented in hardware
and utilizes partial detector information to accept events at a rate of
100~kHz. The high-level trigger is based on software and reduces the rate of
accepted events to 1~kHz.
An extensive software suite~\cite{ATL-SOFT-PUB-2021-001} is used in the
reconstruction and analysis of real and simulated data, in detector operations,
and in the trigger and data-acquisition systems of the experiment.

\section{Data and MC event samples}
\label{sec:samples}
 
The data used in this measurement were collected during the LHC 2015 and 2016
$pp$-collision runs at $\sqrt{s} = 13~\TeV{}$, corresponding to an integrated
luminosity of 36.1~fb$^{-1}$.  The uncertainty in the combined 2015--2016
integrated luminosity is 2.1\%~\cite{ATLAS-CONF-2019-021}, obtained using the
LUCID-2 detector~\cite{Avoni:2018iuv}
for the primary luminosity measurements.
Analyzed events were required to have all ATLAS subdetectors fully operational,
and stable beam conditions.
 
MC-simulated event samples were used in this analysis to
estimate the contamination from background processes, correct the data
from reconstruction level to particle level (unfolding), and for comparisons to the
unfolded data. Four processes are considered in detail for this analysis:
production of \zjets, \wjets, \ttbar, and electroweak diboson events. Single top-quark
production was shown to make a negligible contribution, as its acceptance is similar to
the already small \wjets contribution, and its cross-section is orders of magnitude smaller.
QCD multijet events were shown to be negligible using a data-driven method.
 
The \zjets signal and \wjets background were simulated at next-to-leading-order (NLO) accuracy with the
\sherpa~2.2.1~\cite{Bothmann:2019yzt,Gleisberg:2008ta} MC generator, matching
additional hard parton emissions~\cite{Gleisberg:2008fv} to the parton shower
algorithm based on the Catani--Seymour dipole
formalism~\cite{Schumann:2007mg}. The \mepsatnlo
prescription~\cite{Catani:2001cc,Hoeche:2009rj,Hoeche:2012yf,Hoeche:2011fd} was
used with a merging threshold of 20~\GeV{} to provide merged matrix-element and
parton-shower calculations accurate at NLO in QCD for up to two additional
partons and accurate at LO for up to four additional partons; the virtual QCD
matrix-element components at NLO accuracy were provided by the \openloops
library~\cite{Cascioli:2011va,Denner:2016kdg}. In this configuration,
$g \to \bbbar$ splittings can originate either from the matrix element or from
the parton shower, depending on the transverse scale of the splitting
relative to the CKKW merging scale~\cite{Hoeche:2009rj,Hoche:2010kg}.  As the
matrix-element portions of this calculation include $g \to \bbbar$ splittings,
it is expected that the majority of boosted jets in this sample will have been
initiated by matrix-element rather than parton-shower splittings, but that the
\bbbar splitting within them can arise from either matrix-element or
parton-shower modeling. The 5FNS \textsc{NNPDF3.0nnlo} PDF set~\cite{Ball:2014uwa} with
$\alphas(m_\Z) = 0.118$ was used, in conjunction with the \sherpa authors' standard set of tuned MC
parameter values, referred to as the ``tune''. The samples were normalized to the NNLO inclusive $\Z/W$
cross-sections~\cite{Anastasiou:2003ds}.
 
An alternative LO QCD \zjets sample was simulated using
\mg~2.2.2~\cite{Alwall:2014hca} with up to four additional partons at
matrix-element level, and using the \textsc{NNPDF2.3lo} set of
PDFs~\cite{Ball:2012cx}. This was interfaced with
\PYTHIAV{8.186}~\cite{Sjostrand:2007gs} for modeling of the parton shower and
underlying event, with use of the CKKW-L merging
procedure~\cite{Lonnblad:2001iq,Lonnblad:2011xx}, and with bottom and charm
hadron decays corrected by \evtgen~1.2.0~\cite{EvtGen}.  The A14
tune~\cite{ATL-PHYS-PUB-2014-021} and the 5FNS \textsc{NNPDF2.3lo} PDF
set~\cite{Ball:2012cx} with $\alphas = 0.13$ were used by \PYTHIAV{8}. This
sample was also normalized to the NNLO inclusive cross-section for use in MC-based
background estimation.
 
The \ttbar background was simulated using the \powhegbox~v2
\textsc{HVQ}~\cite{Frixione:2007nw,Nason:2004rx,Frixione:2007vw,Alioli:2010xd}
generator at NLO with the
CT10 PDF~\cite{Pumplin:2002vw}, and matched to the
\PYTHIAV{8.186}~\cite{Sjostrand:2007gs} parton shower and hadronization with the
A14 tune~\cite{ATL-PHYS-PUB-2014-021}.  The top-quark mass was set to 172.5~\GeV, and
the \hdamp parameter -- which controls the \pt of the first additional emission
beyond the Born configuration -- was set to the mass of the top quark. This
sample was normalized to the \ttbar NNLO+NNLL cross-section~\cite{Czakon:2013goa}.
 
The diboson processes ($\W\W$, $\W\Z$, and $\Z\Z$, with one of the bosons
decaying hadronically and the other leptonically) were simulated using
\sherpa~2.1.1 with a \mepsatnlo configuration similar to that used in the
\zjets and \wjets processes described above.  The \textsc{CT10nlo} PDF
set~\cite{PhysRevD.82.074024} with $\alphas = 0.118$ was used, with
the corresponding \sherpa parton-shower tune.
 
All these MC event samples were processed through the ATLAS $\GEANT4$-based
detector simulation~\cite{Agostinelli:2002hh,SOFT-2010-01} and digitization
system to produce inputs to object reconstruction equivalent to those from the
detector data-stream in collision events. Pile-up --- multiple $pp$ collisions in each hard-interaction bunch-crossing, as
well as detector-response effects due to surrounding bunch-crossings ---
was emulated by pre-digitization overlay of simulated detector hits from
multiple \PYTHIAV{8.186} inclusive QCD
events using the A3 tune~\cite{ATL-PHYS-PUB-2016-017}.
The composite events were reweighted in the analysis so the distribution of the
number of overlays per simulated signal event matched the mean number of collisions per
bunch-crossing, $\langle \mu \rangle$, in data.
 
In addition to these MC generator versions and configurations, signal samples
were also produced only at particle level, using the newer \sherpa~2.2.10 generator
in a configuration equivalent to that described above, as well as 4FNS and fusing
variations, and the NLO \mgamc~2.7.3~+~\PYTHIAV{8.244} generator with FxFx
merging in 5FNS and 4FNS modes.
These variations are used in Section~\ref{sec:results} for comparisons with unfolded observables in data.

\section{Lepton and jet definitions}
\label{sec:objectdefn}
 
The objects used in this analysis to select events and define
observables are charged leptons, large-$R$ jets, and (optionally
$b$-tagged) small-$R$ track-jets. These are defined in this section, with
detailed discussion of the systematic uncertainties associated with their
reconstruction postponed to Section~\ref{sec:systematics}.
 
The final results of this analysis consist of observables measured in
particle-level fiducial volumes, which closely match the reconstruction-level event-
and object-selection, to minimize model-dependent extrapolations. In what
follows, the physics objects are defined at both reconstruction level and
particle level. Throughout this analysis, stable particles are defined to be
those with a mean lifetime $\tau > 10~\text{mm}/c$.
 
\begin{description}
\item[Charged leptons:] The leptonically decaying \Z boson is identified by use of high-\pt
charged $e^\pm$ and $\mu^\pm$ pairs, including those from $\tau$-lepton decays.
 
At reconstruction level, identified electrons and muons
were used, with contamination
suppressed by use of ``tight'' and ``medium'' identification criteria for electrons
and muons respectively~\cite{PERF-2017-01,PERF-2014-05}. The lepton candidates were
geometrically restricted to the active regions of the calorimeters and muon spectometer
($|\eta| < 2.47$ excluding the $1.37$--$1.52$ region for electrons, and
$|\eta| < 2.47$ for muons), and have $\pt > 27~\GeV$ for both lepton flavors. Both the
electrons and muons were required to be isolated from significant
energy deposits in the calorimeter and from high-momentum tracks.
 
Corrections derived in $\Z\rightarrow \ell\ell$ events were applied to account
for differences in reconstruction and identification efficiencies between data
and simulated events. Electron energies as measured by the electromagnetic
calorimeter were calibrated to the
electron energy in simulation, and discrepancies between data and simulation
corrected~\cite{PERF-2017-03}. The reconstructed muon momenta were similarly
calibrated using a mix of simulation and data-driven
methods~\cite{ATLAS:2020auj}. Uncertainties associated with the lepton
efficiencies, scales, and resolutions were propagated via systematic
variations.
 
At particle level, the charged leptons are defined as final-state electrons
and muons dressed with direct photons within a surrounding cone of size
$\Delta{R} = 0.1$, with kinematic requirements of $|\eta| < 2.47$ and
$\pt > 27~\GeV$ on the resulting objects.  No explicit requirements of
final-state isolation nor direct connection to the hard scattering are made in
the fiducial lepton definition, hence events with two high-\pt electrons or
two muons in the final-state acceptance are treated as part of the signal even
if one or both arise from $\tau$-lepton or heavy-flavor hadron
decays.
 
\item[Large-$R$ jets:] A \ljet is required in this analysis as a proxy for a
high-momentum, hadronically decaying or splitting object, e.g.~a high-energy
gluon.
 
In reconstructed events, large-$R$ jets were reconstructed from calibrated
topological clusters of calorimeter cells~\cite{Aad2017}, using the anti-$k_t$
algorithm with radius parameter $R=1.0$~\cite{Cacciari:2008gp,Fastjet}. The
clustered jet's energy and pseudorapidity were further calibrated using
simulated data, and its mass was calibrated using a combination of calorimeter
and tracking information~\cite{JETM-2018-02,JETM-2018-06}. Pile-up and
underlying-event contributions to the jets were suppressed by a dynamic
trimming~\cite{Krohn:2009th} procedure discarding clusters from $R=0.2$
subjets with less than 5\% of the original jet \pt. The trimmed jets were
required to have $\pt > 200~\GeV$ and $|\eta| < 2$, to
ensure that the majority of the jet lies within the tracker volume.
Discrepancies between data and simulation in the jet calibration were treated
as systematic uncertainties.
 
At particle level, all final-state particles are used as inputs to the
anti-$k_t$ $R=1.0$ jet algorithm, and trimming is applied with the same
parameters as in reconstruction, effectively subtracting underlying-event
contributions. Again the trimmed jets are required to have $\pt > 200~\GeV$
and $|\eta| < 2$.
 
\item[Subjets and $b$-tagging:] Small-$R$ subjets within the \ljet are
used as proxies for the leading partons in the jet, e.g.~the $b$-quarks in a
high-energy $g \to \bbbar$ splitting.  To achieve high angular
resolution,
inner-detector tracks are used in place of calorimeter
information to construct these subjets.
 
At reconstruction level, the anti-$k_t$ algorithm with radius parameter $R=0.2$
was used to construct track-jets from at least two inner-detector tracks matched to the primary vertex (see
Section~\ref{sec:eventseln}). The track-jets were required to have $\pt > 10~\GeV$ and
$|\eta | < 2.5$. Identification of the track-jets as likely (or not) to have
been initiated by $b$-quarks was provided by the ATLAS MV2c10 multivariate
$b$-tagging algorithm~\cite{JETM-2018-03}
trained on leptonic \ttbar events to achieve 70\% $b$-tagging efficiency,
with mis-tag rejection ratios of
7.1 and 120 for charmed and light jets respectively.
Corrections were applied to compensate for
differences in $b$-tagging efficiency and charm- and light-jet mis-tag rates
observed between simulation and collision data. Systematic uncertainties from
tracking, vertexing, and $b$-tagging calibration were evaluated.
 
At particle level, stable charged particles are used as the inputs to the
$R=0.2$ anti-$k_t$ jet algorithm, again with $\pt > 10~\GeV$ and
$|\eta | < 2.5$. A charged-particle jet is considered to be $b$-tagged if a
weakly decaying $b$-hadron, with $\pt > 5~\GeV$, is associated with it by the
ghost-association method~\cite{Cacciari:2008gn}.
 
Any small-$R$ track-jets or small-$R$ charged-particle jets matched to the
large-$R$ jets by ghost-association are considered to be charged subjets of
the large-$R$ jet. A large-$R$ jet is considered $b$-tagged if any of its
small-$R$ subjets is $b$-tagged, and the number of $b$-tagged subjets is used
to define subclasses of signal events.
\end{description}
 
A simple overlap-removal procedure was used at both reconstruction and particle
levels to accommodate a single particle or object leaving multiple signatures in
the detector. In particular, this procedure was motivated by the possibility
that leptons from the \Z-boson decay could also be recorded as an additional
large-$R$ jet. To correct for this, the angular separation between each lepton
and large-$R$ jet was computed, and if any were within $\dr = 1.0$ of each other
then the jet was removed. This cut has the additional effect of suppressing
contributions from the dijet process with collinear \Z-boson emission from
quarks, while still admitting widely separated topologies in which the \Z boson
and the large-$R$ jet are located in the same event hemisphere.

\section{Event selection and observables}
\label{sec:eventseln}
 
For both the data and reconstruction-level MC samples, single-electron and
single-muon triggers~\cite{ATLAS:2019dpa,ATLAS:2020gty,ATLAS:2021lws} were used to select the
subset of events of \emph{a priori} relevance to this analysis before application of offline
event-selection cuts.
The kinematic requirements on the leptons defined in the previous section ensured full
trigger-efficiency for events in the analysis fiducial phase-space. Candidate
events were required to have a primary vertex, defined as the vertex with the
highest sum of track-$\pt^2$, and with at least two associated tracks of
$\pt > 400~\MeV$.
 
Event selection cuts on the physics objects defined in
Section~\ref{sec:objectdefn} were applied equally to reconstruction-level
events in data and MC simulation, and to particle-level MC events. To select signal events
containing a leptonically decaying \Z-boson candidate produced in association
with a \ljet, events were required to have exactly two charged leptons of the
same flavor, $\ell^\pm \in \{ e^\pm, \mu^\pm \}$, and at least one \ljet.  No
opposite-charge requirement was placed on the charged-lepton pairs, but the
invariant mass of the lepton pair, $m_{\ell \ell}$, was required to be greater
than 50~\GeV{} to exclude the photon-dominated part of the Drell--Yan continuum.
 
This set of criteria defines the ``inclusive'' event-selection region
for this analysis. A more exclusive ``2-tag'' region was defined as a subset of
this by additionally requiring that the \ljet contains exactly two $b$-tagged
subjets. The numbers of events selected in these regions, for the dielectron
and dimuon channels separately, and for both collision data and MC simulation,
are shown in Table~\ref{tab:yields}, omitting the single-top and multijet
processes, which were shown to be negligible by MC studies and a data-driven
background study respectively.
 
The \ljet used to construct the observables below was chosen as the highest-\pt
\ljet in the inclusive selection, and the highest-\pt 2-tag \ljet in the 2-tag
selection. The 2-tag observables are hence not a subset of the inclusive
observables, as the latter includes events in which the highest-\pt large-$R$ jet
does not contain two $b$-tagged subjets.
 
\begin{table}
\centering
\caption{Reconstruction-level event-selection yields (and statistical
uncertainties of the expected yields) in the $ee$ and $\mu\mu$ channels from
each process's MC sample (with \sherpa~2.2.1 used for the \zjets samples)
with the normalizations discussed in Section~\ref{sec:samples}, and from
collision data. 
The $\Z$+\qqbar categories were defined using a particle-level filtering strategy
in which the given orthogonal combinations of heavy hadrons (from any source)
with $\pt > 5~\GeV$, $|\eta| < 2.9$ and associated to a $\pt > 10~\GeV$ truth-particle
jet, were enforced in separate MC event samples at during event generation. 
The single-top
process was found to make a negligible contribution to all event selections
and has been omitted. Multijet backgrounds were estimated to be negligible
by a data-driven method.}
\vspace*{1ex} \sisetup{separate-uncertainty=true,round-precision=1} \small
\begin{tabular}{llS[table-format=5.0(3)]S[table-format=5.0(3)]cS[table-format=3.1(2)]S[table-format=3.1(2)]}
\toprule
& \hspace*{1em} & \multicolumn{2}{c}{Inclusive} & \hspace*{1em} & \multicolumn{2}{c}{2-tag}\\
\cmidrule{3-4} \cmidrule{6-7}
& & \multicolumn{1}{r}{$ee$} & \multicolumn{1}{r}{$\mu\mu$} & & \multicolumn{1}{r}{$ee$} & \multicolumn{1}{r}{$\mu\mu$}\\
\midrule
$\Z$+\bbbar & &   324 +-   4 &   305 +-   4 & & 163.8 +- 2.6 & 157.2 +- 2.5 \\
$\Z$+\ccbar & &   536 +-  10 &   530 +-   9 & &  12.3 +- 1.8 &  19.3 +- 2.0 \\
$\Z$+$bc$   & &    89 +-   2 &    81 +-   2 & &  14.6 +- 1.2 &  12.1 +- 0.9 \\
$\Z$+$b$    & &  2588 +-  13 &  2423 +-  12 & &  14.8 +- 1.1 &  12.4 +- 1.3 \\
$\Z$+$c$    & &  5073 +-  32 &  4862 +-  39 & &   5.5 +- 1.3 &   6.9 +- 1.7 \\
$\Z$+light  & & 53808 +- 164 & 51206 +- 145 & &   9.4 +- 1.1 &  11.1 +- 1.5 \\
\ttbar      & &  5960 +-  46 &  5204 +-  43 & &  82.7 +- 5.3 &  75.4 +- 5.6 \\
$\W$+jets   & &    73 +-   4 &     7 +-   1 & &   0.4 +- 0.1 &  \multicolumn{1}{l}{$< 0.1$} \\
Diboson     & &  2042 +-  17 &  1834 +-  16 & &  21.5 +- 1.4 &  20.7 +- 1.4 \\
\midrule
MC total    & & 70493 +- 175 & 66452 +- 158 & & 324.9 +- 6.8 & 315.1 +- 7.2 \\
Data        & & 66481        & 65034        & & 391        & 384        \\
\bottomrule
\end{tabular}
\label{tab:yields}
\end{table}

The differential observables measured in this analysis are:
\begin{description}
\item[Properties of the \ljet (\J):] the \ljet mass, \mJ, and transverse momentum,
\ptJ, for both the inclusive and 2-tag regions;
\item[Properties of the \ljet and \Z-boson system (\ZJ):] the transverse momentum of
their vector sum, \ptZJ, and their separation in $\phi$, \dphiZJ, for the
inclusive region only;
\item[Subjet separation:] the angular separation, \drbb, between the two
$b$-tagged subjets 
in the 2-tag region.
\end{description}
These variables respectively measure the external and internal kinematics
of the selected \ljet, the effect of additional QCD radiation on the \ZJ event
topology, and the kinematics of $g \to \bbbar$ splitting in the boosted
regime. The total cross-sections in the fiducial volume for the inclusive and 2-tag event
selections were also measured, via integration of the differential measurements.
 
\begin{figure}
\centering
\subfigure{\includegraphics[width=0.48\textwidth]{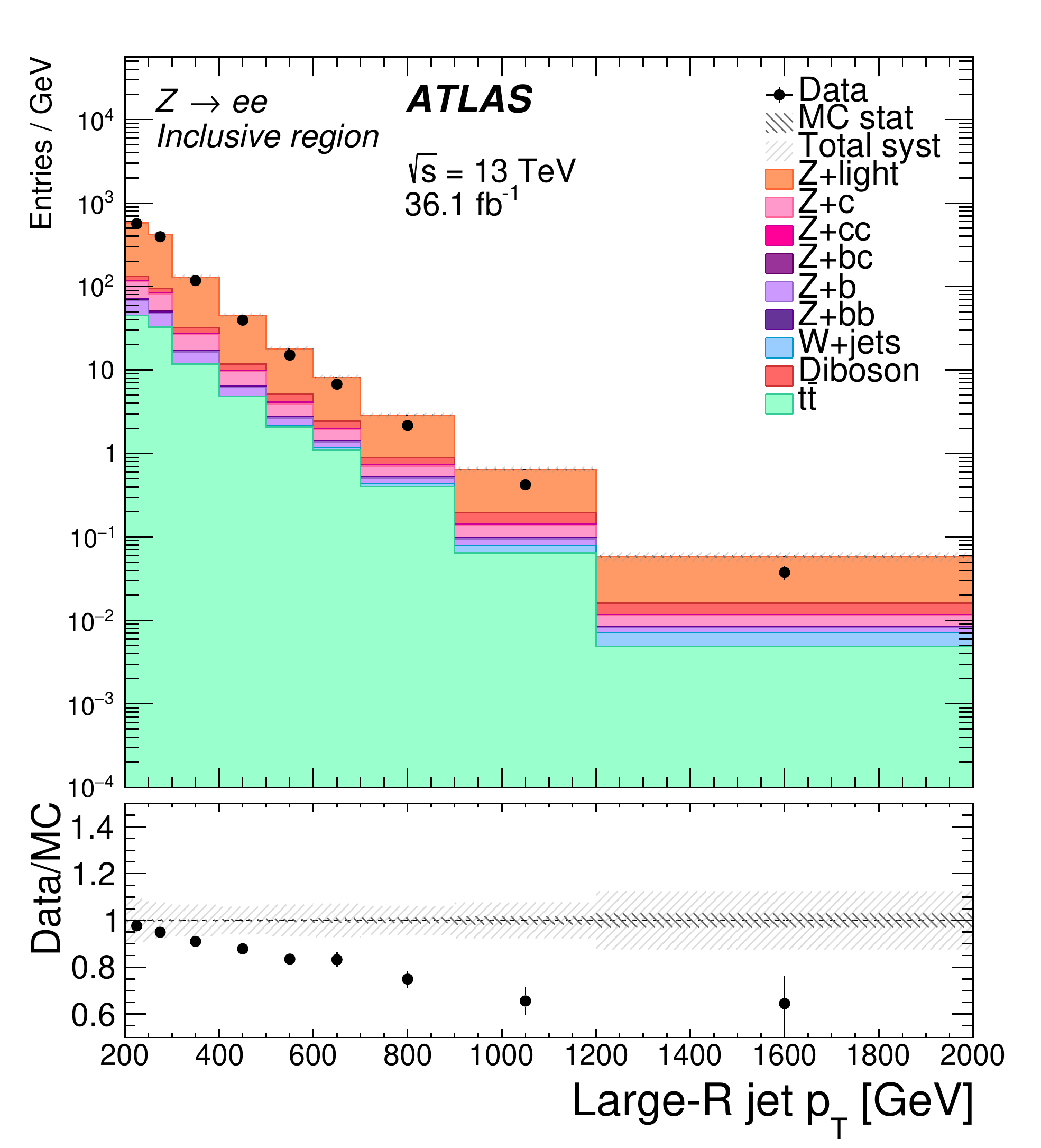}}
\subfigure{\includegraphics[width=0.48\textwidth]{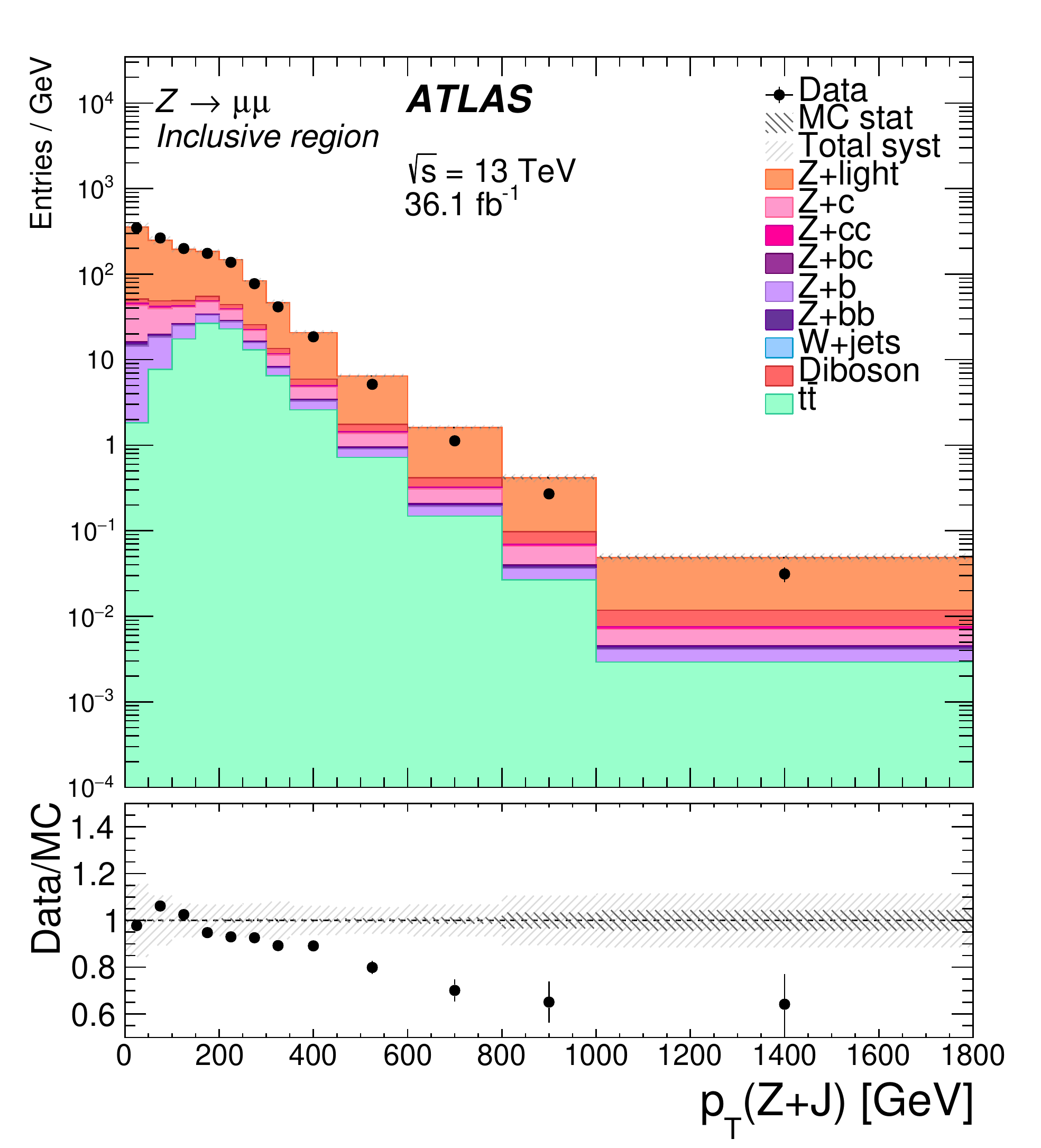}}
\subfigure{\includegraphics[width=0.48\textwidth]{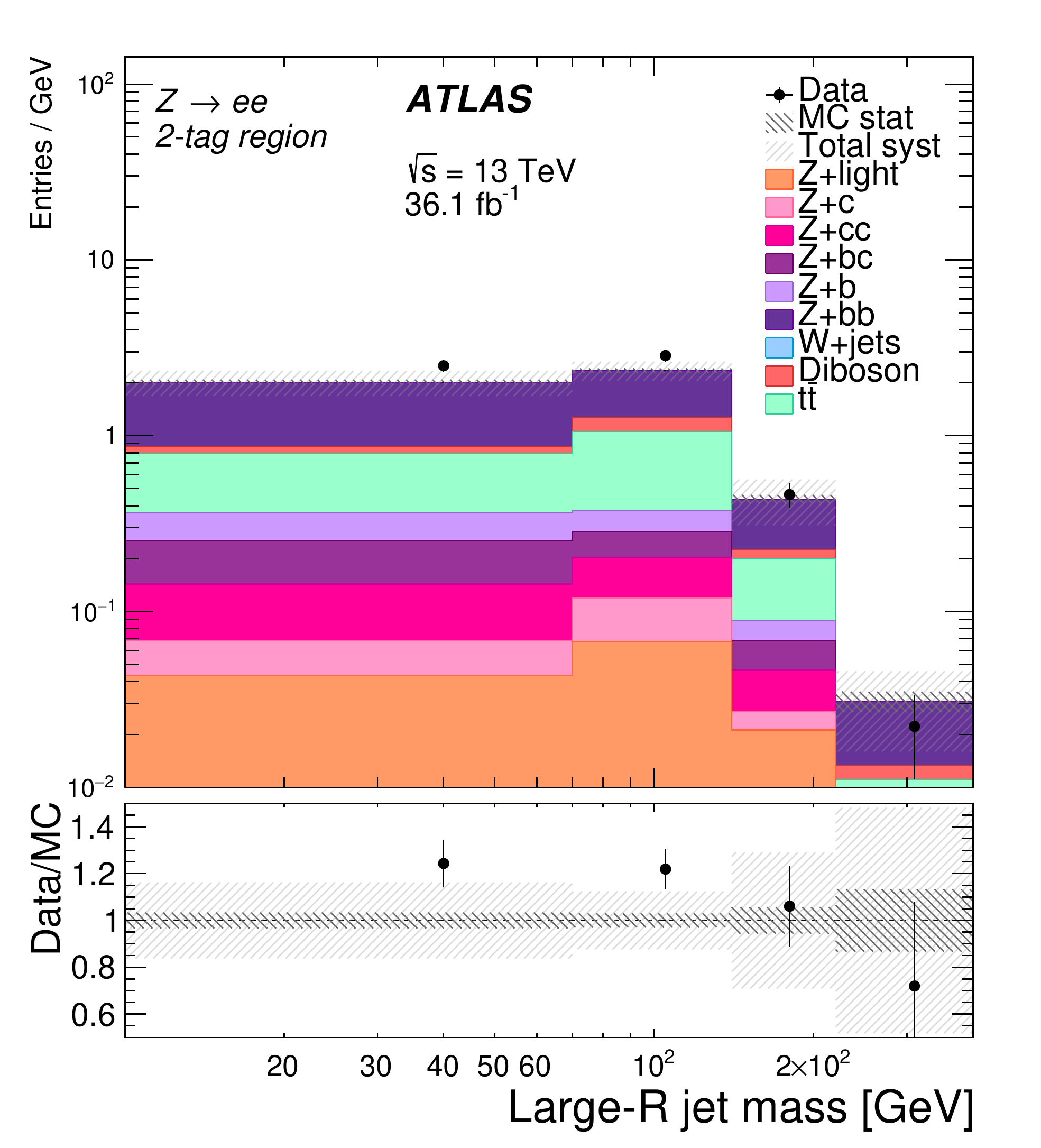}}
\subfigure{\includegraphics[width=0.48\textwidth]{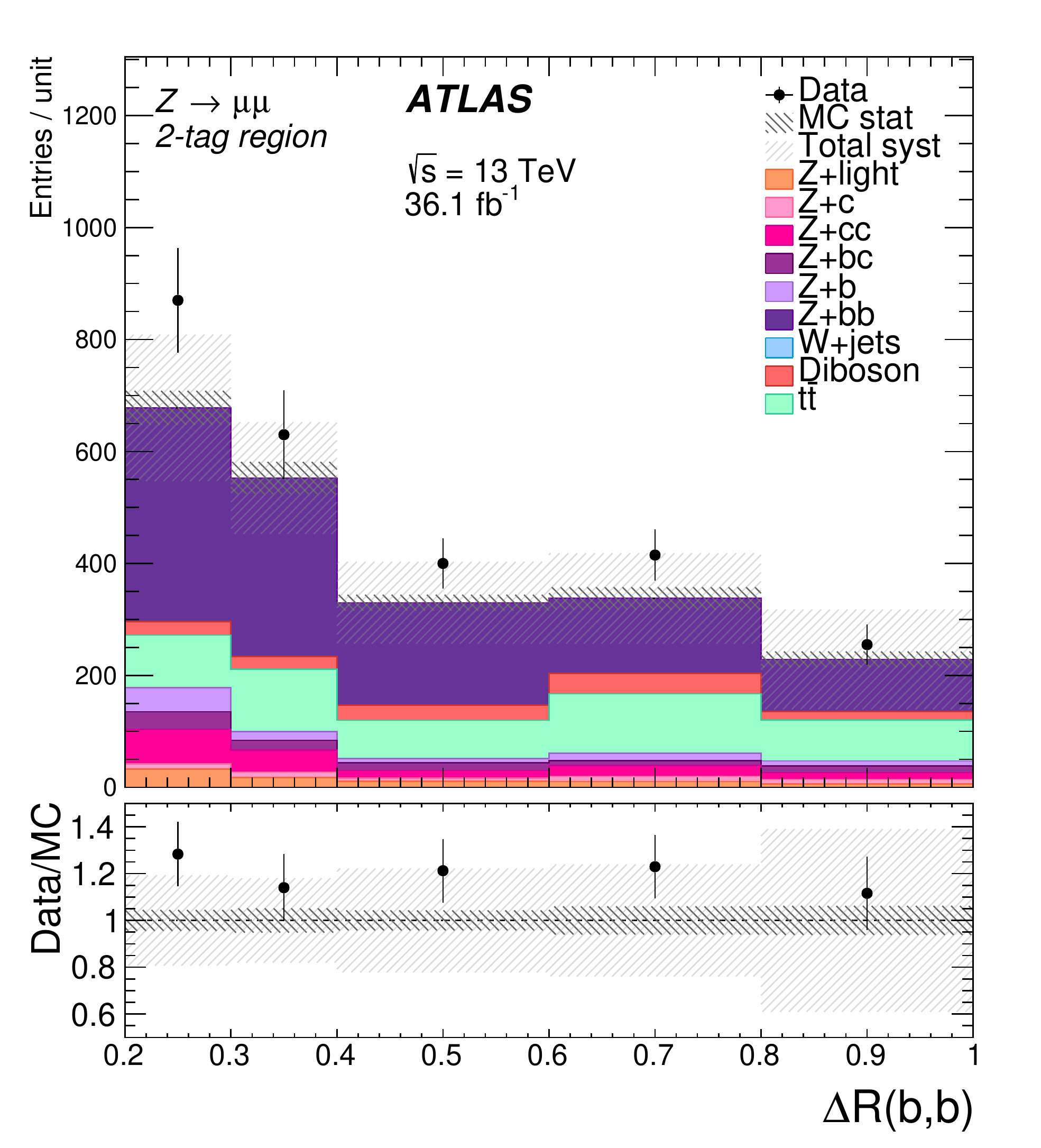}}
\caption{Selected reconstruction-level observables, compared with pre-fit MC
simulation with \sherpa~2.2.1 used for the \zjets samples: the top row shows the inclusive-selection $ee$ \ljet \pt
distribution (left) and $\mu\mu$ \ZJ \pt distribution (right), and the
bottom row shows the 2-tag selection $ee$ \ljet mass and $\mu\mu$ \drbb
distributions. The MC statistical uncertainties are shown by the dark gray
band and the total uncertainty, including in quadrature the systematic
uncertainties detailed in Section~\ref{sec:systematics}, are shown by the
light gray band. The statistical uncertainty of the data is given by the
error bar on the data point.}
\label{fig:recoplots}
\end{figure}

A selection of these observables are shown at reconstruction level in
Figure~\ref{fig:recoplots}, illustrating the inclusive \ljet and \ZJ \pt
distributions, and the 2-tag \ljet mass and \drbb distributions in data and simulation.
One $ee$ and one $\mu\mu$ dilepton observable are shown for each event-selection
region: all observables were constructed separately for the electron and muon
channels, to allow consistency checks and consideration of distinct
detector effects before lepton-channel combination. Significant
discrepancies between data and the sum of MC-modeled processes (including
correction factors) are visible in several variables, particularly the
inclusive \ljet and \ZJ-system transverse momenta. This evidence of general
mismodeling in the boosted phase-space at reconstruction-level
motivated the publication of the detector-corrected forms of these observables.
 
These plots show the admixture of processes contributing to the two
event-selection regions, with the inclusive event-selection dominated by the
\Z{}+light-parton (including \Z{}+\cq) and \ttbar production processes, and the 2-tag region
dominated by \Z{}+\bbbar with significant background contributions from diboson
and \ttbar production.

\section{Correction of observables to particle level}
\label{sec:unfolding}
 
The main result of this analysis is the set of differential cross-sections
introduced in the previous section, corrected to the particle-level fiducial
phase-space by unfolding: the deconvolution of biases introduced by the detector and
reconstruction algorithms.
Presentation in the fiducial form assists comparison with results from
different experiments and with theoretical predictions.
 
In this analysis, the unfolding was performed using the Fully Bayesian
Unfolding~(FBU) technique~\cite{FBU}. FBU directly performs a likelihood fit in
the parameter space of signal cross-sections $\vec{\sigma}$, plus a set of
nuisance parameters $\vec{\nu}$ that control background compositions and
systematic uncertainties (to be described in Section~\ref{sec:systematics}). The
FBU method hence gives access to a full posterior probability density in the
space of signal and nuisance parameters, from which arbitrarily detailed
correlation information may be extracted.
 
The FBU posterior probability for each observable is constructed as the product
of Poisson probabilities over all reconstruction-level bins $i$ as a function of
the model parameters $\vec{\theta} = \{ \vec{\sigma}, \vec{\nu} \}$,
\begin{equation}\nonumber
\mathcal{P}(\vec{\theta} | \vec{d})
\propto
\llh(\vec{d} | \vec{\theta}) \cdot
\pi(\vec{\theta}) \, ,
\end{equation}
where $\vec{d}$ is the set of observed bin counts in data, and $\pi(\vec{\theta})$ is
a set of prior probability densities over the model parameters. The \llh term in this can be
expressed
as a product of Poisson likelihoods over the bins,
\begin{equation}\nonumber
\llh(\vec{d} | \vec{\theta}) =
\prod_{i \in \, \text{bins}}
\mathrm{Poiss}(d_i, x_i(\vec{\theta})) \, ,
\end{equation}
where $\vec{x}(\vec{\theta})$ is the set of expected total bin yields. This can
be decomposed further into reconstruction-level background and signal cross-sections $b_i$ and $s_i$,
\begin{align}
x_i(\vec{\theta})
&= \lumi(\vec{\nu}) \cdot \big( b_i(\vec{\nu}) + s_i(\vec{\theta}) \big) \nonumber \\
&= \lumi(\vec{\nu}) \cdot \Big( b_i(\vec{\nu}) + \sum_j \respm_{ij}(\vec{\nu}) \cdot \sigma_{\!j} \Big) \, , \nonumber
\end{align}
where $\lumi$ is the integrated luminosity of the dataset and $\respm_{ij}$ is
the response matrix $p(i|j)$ mapping particle-level
bins $\{j\}$ to reconstruction-level bins $\{i\}$. Overflows outside the
fiducial acceptance are included in the bin indexing, so migrations into and out
of the acceptance are treated with the full machinery.  This provides the full
formalism necessary to relate the observed data $d_i$ to our parameters of
interest, the particle-level signal cross-sections $\sigma_i$.
In the inclusive region, the signal is defined as all \zjets contributions,
while in the 2-tag region it is only \Z{}+\bbbar, with other \zjets flavors now
considered as part of the background.
 
\begin{figure}
\centering
\subfigure{\includegraphics[width=0.48\textwidth]{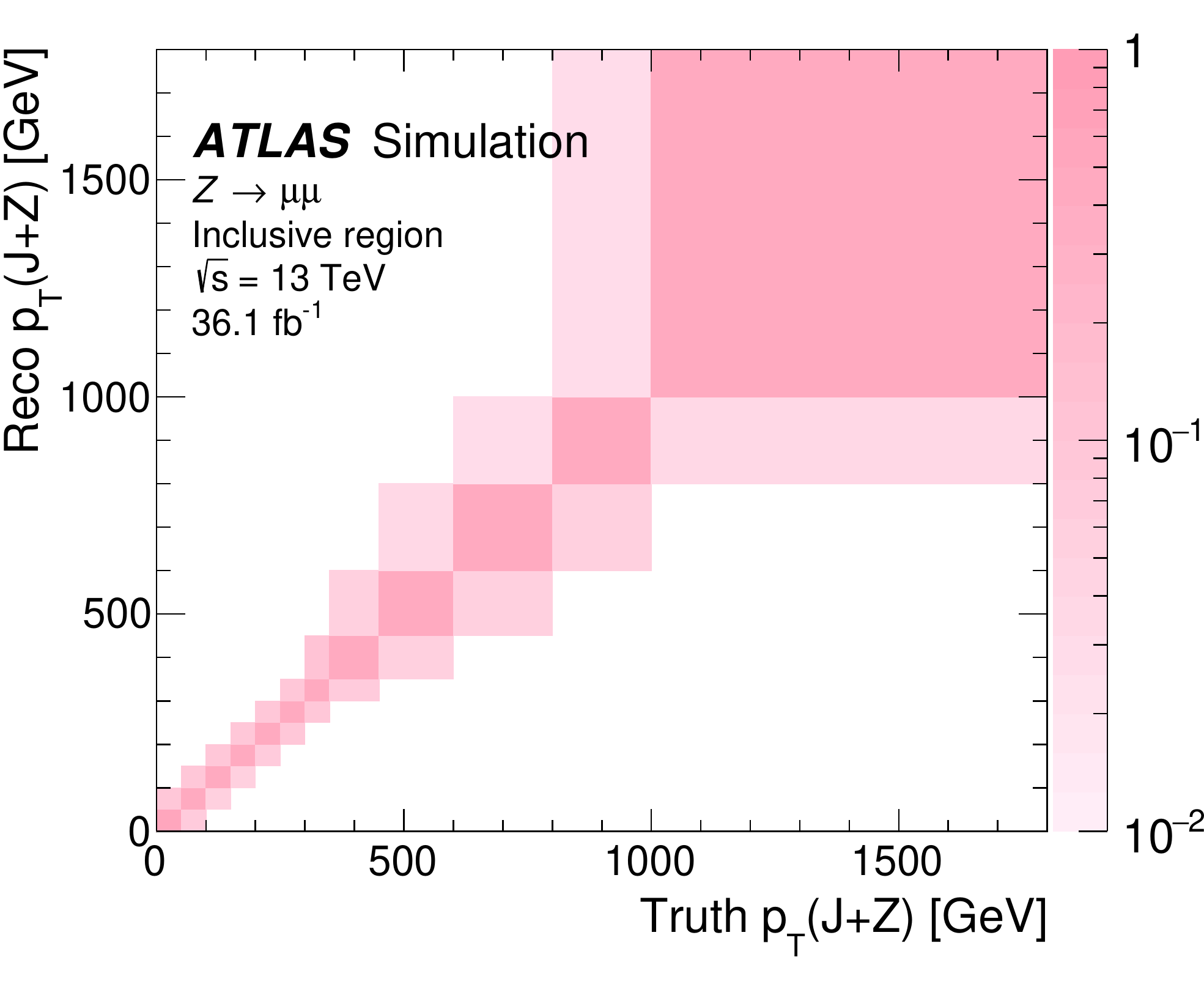}}
\subfigure{\includegraphics[width=0.48\textwidth]{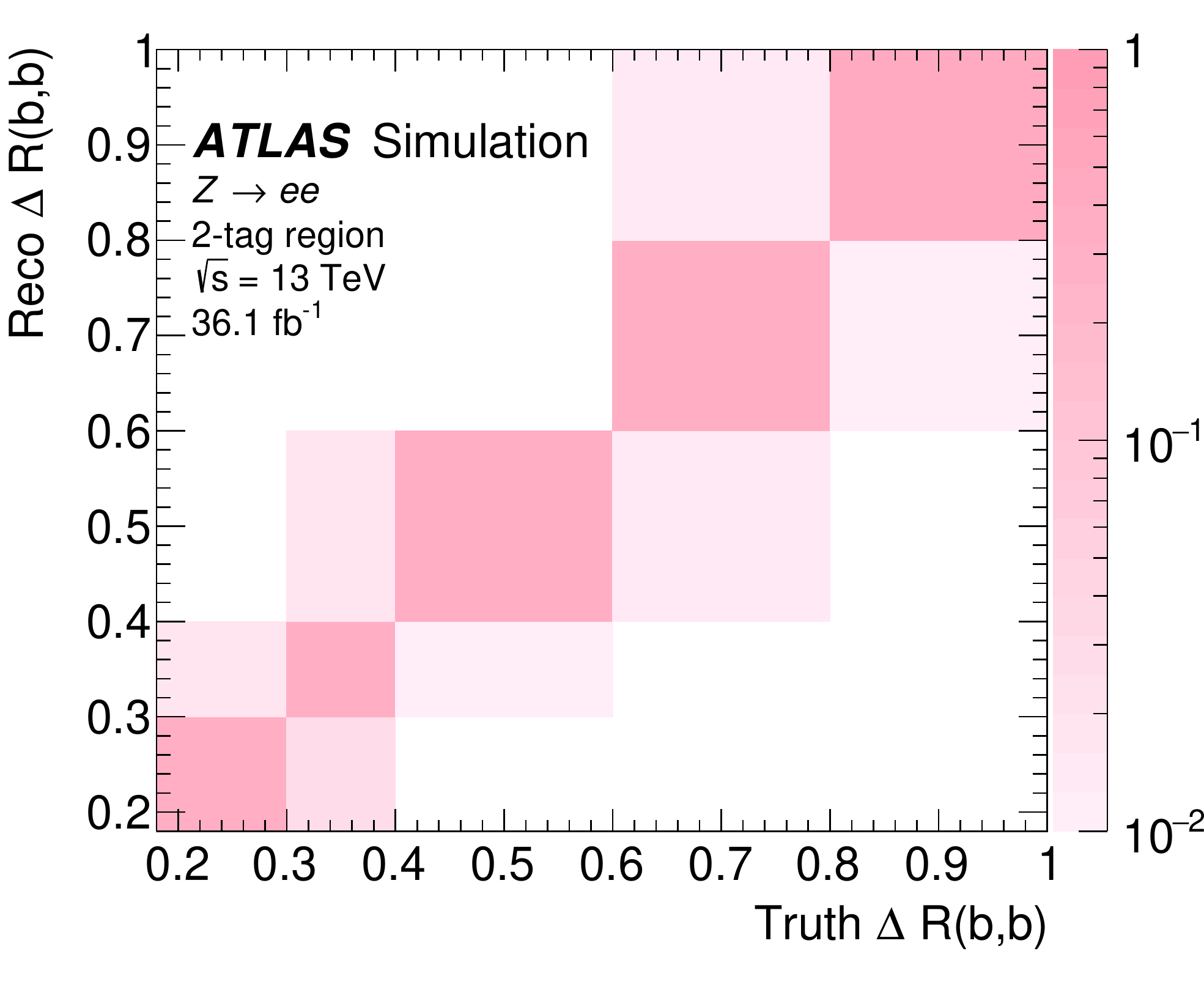}}
\caption{Example response matrices for inclusive-region $\mu\mu$-channel \ljet
+ \Z \pt (left) and 2-tag $ee$-channel \drbb (right), with nominal signal
modeling.}
\label{fig:responses}
\end{figure}
 
The background cross-sections and response matrix were constructed from a set of
MC-derived histogram templates, including predictions from the nominal MC samples
(\sherpa \zjets) with data/MC corrections, and a set of predictions
from each systematic-uncertainty variation to be described in
Section~\ref{sec:systematics}.
Examples of nominal-model response-matrix templates for two observables are
shown in Figure~\ref{fig:responses}.
 
The unit nuisance parameters $\nu_k \in \vec{\nu}$ were used to define linear
interpolations of $b_i$ and $\respm_{ij}$ between templates corresponding to
$\nu_k = 0$ and~$1$. The absolute deviations obtained from ``up'' and ``down''
variations of nuisance parameters were averaged into single positive deviations
for use in this symmetrized form. Unit Gaussian priors were applied to all
$\nu_k$ other than the luminosity uncertainty; as negative luminosities would
imply unphysical negative event rates, the luminosity uncertainty was modelled
by an always-positive log-normal prior with $\mu = 0$ and $\sigma = 0.021$.
The background normalizations were allowed to float with Gaussian prior widths discussed
in Section~\ref{sec:systematics}, and flat, non-negative priors were imposed on
the signal cross-section parameters $\vec{\sigma}$.
 
For this analysis, the ``hunfold''~\cite{hunfold} implementation of FBU was
used. This uses gradient ascent to maximize the posterior log-probability
$\ln \mathcal{P}(\vec{\theta})$, and then samples the posterior probability
distribution using a proposal density in $\vec{\theta}$ derived from the
likelihood Hessian matrix at the maximum-likelihood point. In the unfolded
observables of this analysis (see Section~\ref{sec:results}), each variable is
constructed from a sum of electron and muon channels via a single FBU fit, with
a ``double width'' concatenation of electron and muon response matrices used to
simultaneously unfold the electron and muon distributions into combined
particle-level distributions representing \Z{}-boson decays into either lepton flavor.

\section{Systematic uncertainties}
\label{sec:systematics}
 
The measurements in this analysis are affected by statistical uncertainties and
by systematic uncertainties from detector-interaction and reconstruction
processes, from MC modeling, and from the unfolding procedure. Estimates of
these uncertainties were derived using standard methods described in this
section, and were propagated through the unfolding procedure
where they affected the final posterior distributions.
 
The main sources of experimental
systematic uncertainty affecting these measurements were:
\begin{description}
\item[Charged leptons:] energy/momentum scale and resolution, and reconstruction,
identification, isolation, and trigger efficiencies. Systematic variations of
the data/MC efficiency corrections and energy/momentum calibrations applied to
the MC samples~\cite{PERF-2017-03,PERF-2011-01} were used to define variations from their nominal templates in
the FBU unfolding's parameterized background cross-sections and signal response
matrices.
 
\item[Large-$R$ jets:] energy scale, mass scale, energy resolution, and mass
resolution. The jet energy scale (JES) and jet mass scale (JMS) uncertainties
were based on the double-ratio of each variable (energy, mass) to its
equivalent reconstructed from track-jets, between data and simulation; this
construction permitted separation of the physics effects from the calorimeter
reconstruction systematic uncertainties~\cite{ATL-PHYS-PUB-2015-015}. In this analysis, the
JES and JMS uncertainties have been treated as fully uncorrelated;
cross-checks assuming higher degrees of correlation had negligible effect.  The
uncertainty in the jet mass resolution (JMR) was determined by smearing the
jet mass such that its resolution was degraded by 20\%, and for the \ljet
energy resolution (JER), 
symmetric variations of the jet energies by $\pm 2\%$ were applied.
 
\item[Flavor tagging:] track-jet $b$-tagging efficiencies, and mis-tag rates
for $c$-jets and light-flavor jets.
The $b$-tagging efficiency and charm mis-tag rate in simulation were calibrated
using the tag-and-probe method in \ttbar events~\cite{PERF-2016-05,ATLAS-CONF-2018-001},
and the light-jet mis-tag rate was calibrated in dijet events~\cite{ATLAS-CONF-2018-006}.
A total of~25 diagonalized systematic uncertainties associated with these
calibration factors were considered in this analysis.
 
\item[Pile-up:] pile-up reweighting uncertainty. The MC predictions were
reweighted such that their distribution of the number of pile-up vertices matched
the pile-up distribution measured in the data. The uncertainty from this
procedure was propagated to the unfolding using variations of pile-up weights
to account for the uncertainty in the pile-up estimation.
\end{description}
 
In addition to the above measures of imperfect understanding of the detector, uncertainties in the
modeling of both the signal and background physics processes were propagated to the
final measurement, via systematic variations in the unfolding components.  As for
the detector uncertainties, this propagation was implemented via linear interpolation of
template distributions (including response matrices) in the unfolding machinery.
For the signal process, standard 7-point variations of the renormalization and factorization
scales by factors of two from the nominal values, the nominal PDF's error-set,
differences between the nominal PDF and the alternative
\textsc{CT14nnlo}~\cite{Dulat:2015mca} and \textsc{MMHT2014nnlo}~\cite{Harland-Lang:2014zoa}
central PDFs, and variations of $\alphas(m_Z)$ by~$\pm 0.001$ were included.
Approximate uncertainties in the matching procedure between matrix
element and parton shower were evaluated by variations between the nominal
\sherpa samples and the \mgamc+~\PYTHIAV{8} samples. 
The envelope of differences in the signal-process response matrix between
\sherpa and \mg was also treated as a systematic uncertainty; a
cross-check showed that the results were not sensitive to whether or not the
\sherpa/\mg difference was split into several independently parametrized
components.
 
Dedicated variation MC samples were used to evaluate the modeling of the \ttbar
background, considering the matrix element, parton shower model, and the
dependence on initial- and final-state radiation settings and the \hdamp
parameter.
In addition, the quality of \ttbar modeling was assessed in MC--data comparisons
using an opposite-flavor $e \mu$ variation on the standard $ee$/$\mu\mu$
selection; additional mismodeling uncertainties of 30--50\%
were added to the first bins of the inclusive \ptZJ observable, to cover an
MC--data discrepancy in this control region.
A conservative $\sigma = 0.2$ Gaussian prior,
informed by the maximum reconstruction-level data/MC disagreement,
was used for the normalization uncertainty of all background processes
other than the non-signal \Z{}+jets samples in the 2-tag region, for which
a larger $\sigma = 0.5$ prior was used.
This inflated uncertainty was assigned to reflect that a robust \emph{in situ}
measure of the 2-tag flavor fractions could not be obtained from $b$-tagging
variable templates due to the low event-count in this analysis phase-space.
 
In addition, systematic errors arise from the finite sizes of the simulation samples.
In principle, the statistical limitation of each bin corresponds to a nuisance
parameter in the unfolding. But in practice many of these parameters have
little effect, since most bins with low rates, e.g.~very statistically
limited off-diagonal elements of the response matrix, by construction do not
contribute much to the result. This abundance of uncertain quantities
creates an intractably large space of nuisance parameters in which the unfolding
fit is unlikely to converge. A ``pruning'' procedure was hence implemented, both
for statistical uncertainties in response-matrix estimates and for all the
detector and modeling systematic uncertainties described. The pruning criterion was to
remove nuisance parameters which produced a background variation of less than
5\% in all bins (the background fractions being around 20\% of the event yield
in the inclusive region and 40\% in the 2-tag region), and did not change any
entries in the response matrix by at least 0.002. The effect was to prune
jet-substructure and most lepton-calibration systematic uncertainties for all
observables, jet-mass systematic uncertainties for the $\ptJ$ and $\ptZJ$ observables, and
jet-mass resolution for all but the \ljet mass observable. The $b$-tagging
uncertainties were partially pruned for the 2-tag region; naturally, there are
no such uncertainties in the inclusive region. The unpruned MC statistical uncertainties in
total contributed a subleading ${\sim}1\%$.
 
\begin{figure}
\centering
\subfigure{\includegraphics[width=0.48\textwidth]{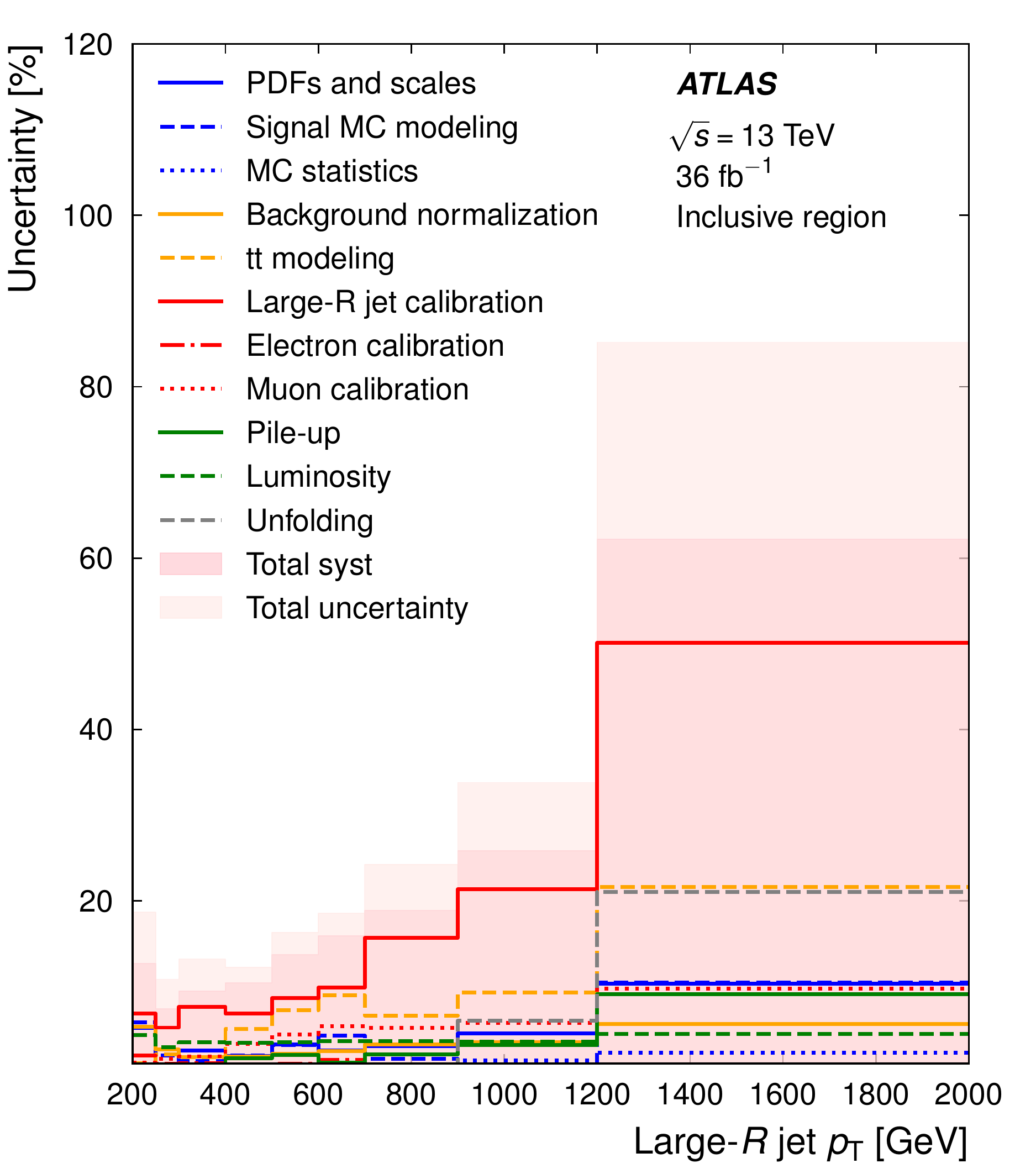}}
\subfigure{\includegraphics[width=0.48\textwidth]{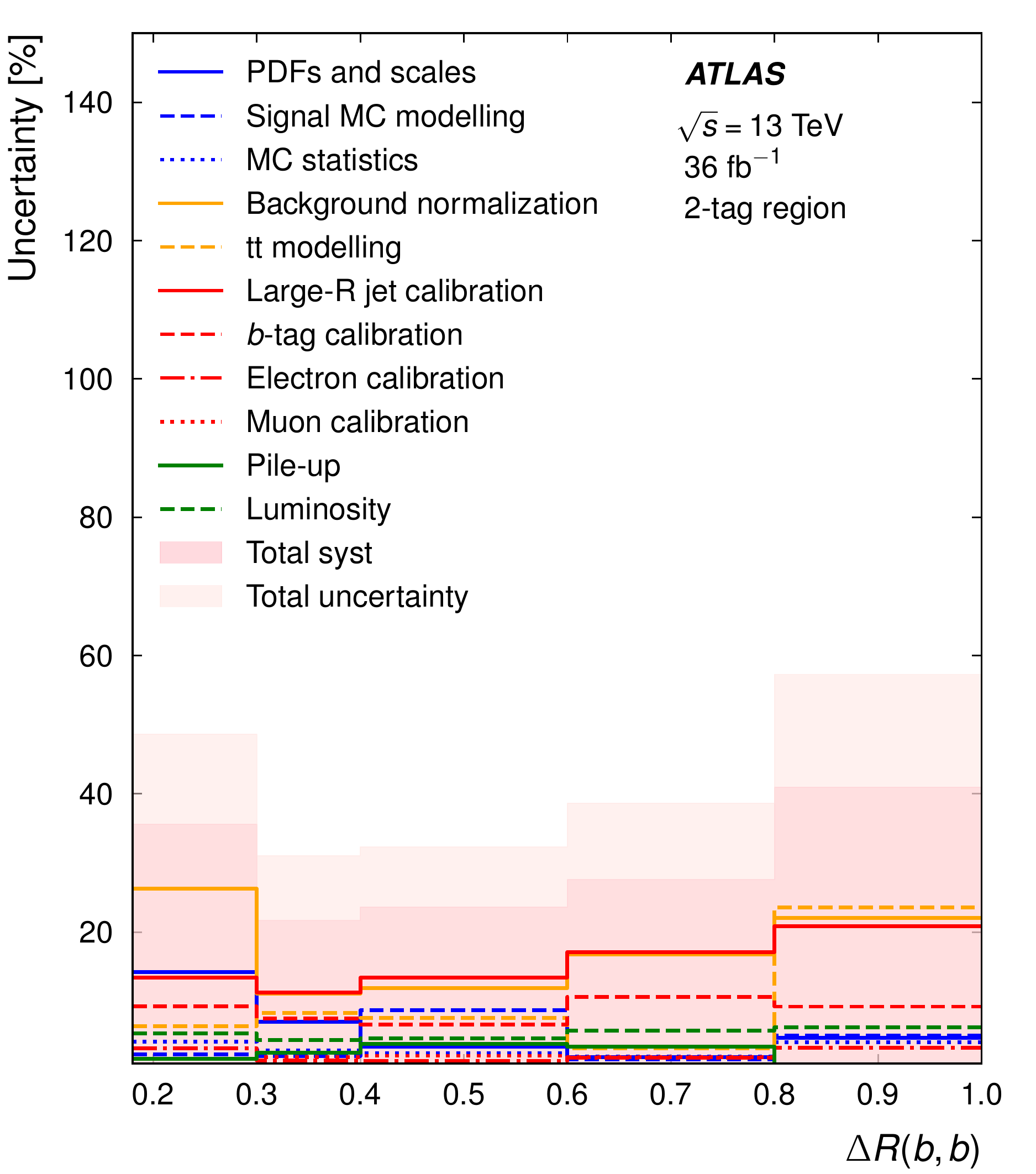}}
\caption{Illustration of leading post-unfolding groups of systematic
uncertainties for the inclusive-selection \ljet \pt (left) and 2-tag
selection \drbb (right). These groups have been constructed from elementary
systematic nuisance parameters, assuming statistical independence of error
sources within each group.}
\label{fig:systsummaries}
\end{figure}
 
A comprehensive set of closure tests in the unfolding\,/\,nuisance-profiling
procedure was performed,
including closure tests with and without reweighting of the MC samples to
match the reconstruction-level data distribution in each variable,
stress-testing by reweighting \sherpa pseudodata with a \mg-derived
response matrix and vice versa, and checks against bias from nonuniform signal
priors. Non-closure effects from these tests, which were very small in nearly
all bins, but rose to around 20\% in single bins of the \ptJ and \ptZJ
distributions, were added in quadrature to the mix of final uncertainties.
 
Summary systematic uncertainties are listed in Table~\ref{tab:systs} and
illustrated for two observables in Figure~\ref{fig:systsummaries}.  The
summaries were obtained from the FBU posterior-distribution samples by computing
the sample covariance matrix between all fit parameters, including nuisance
parameters and signal bin-values,
$\mathrm{cov}_{ij} = \langle x_i x_j \rangle - \langle x_i \rangle \langle x_j
\rangle$ for parameter indices $i$ and $j$. The absolute values of
covariance-matrix rows for the elementary nuisance parameters were then summed
in semantic groupings, e.g.~the sets of nuisance parameters for electron
reconstruction, jet reconstruction, $b$-tagging, etc.
The resulting grouped-covariance entries were projected onto the signal-bin
cross-section parameters via the relation
$\sigma_b = \mathrm{cov}_{bn}/\sqrt{\mathrm{cov}_{nn}}$, where $b$ is the signal
bin index, and $n$ the systematic nuisance index. By construction, these grouped
uncertainties are symmetric. The total uncertainty, including statistical
effects, is given by the larger standard deviation of each signal bin value. The
largest systematic-group effects are from \ljet calibration and signal modeling
in the inclusive region, and from background normalization and $b$-tagging
calibration in the 2-tag region.
 
\begin{table}
\centering
\caption{Summary table of relative uncertainty magnitudes per
observable. These uncorrelated estimates of systematic uncertainties'
contributions to the total uncertainty are based on projection of nuisance
parameters onto the signal cross-section bin values via the likelihood-scan
covariance matrix, and summing the elementary contributions in quadrature.}
\vspace*{1ex}
\small
\footnotesize
\sisetup{round-mode=places,round-precision=0}
\begin{tabular}{l
S@{\!\!--~}S@{\hspace*{-1.1em}\%\hspace*{1.4em}}
S@{\!\!--~}S@{\hspace*{-1.1em}\%\hspace*{1.4em}}
S@{\!\!--~}S@{\hspace*{-1.1em}\%\hspace*{1.4em}}
S@{\!\!--~}S@{\hspace*{-1.1em}\%\hspace*{1.4em}}
c@{\hspace*{1em}}
S@{\!\!--~}S@{\hspace*{-1.1em}\%\hspace*{1.4em}}
S@{\!\!\!\!--~}S@{\hspace*{-1.1em}\%\hspace*{1.4em}}
S@{\!\!\!\!--~}S@{\hspace*{-1.1em}\%}}
\toprule
& \multicolumn{8}{c}{Inclusive}
& \multicolumn{6}{c}{2-tag}\\
\cmidrule{2-9} \cmidrule{11-16}
& \multicolumn{2}{l}{\ptJ}
& \multicolumn{2}{l}{\mJ}
& \multicolumn{2}{l}{\ptZJ}
& \multicolumn{2}{l}{\dphiZJ} &
& \multicolumn{2}{l}{\ptJ}
& \multicolumn{2}{l}{\mJ}
& \multicolumn{2}{l}{\drbb}\\
\midrule
Scales \& PDFs    &  0.4 & 10.3 &  2.5 &  6.7 &  1.8 &  8.5 &  0.5 &  4.0 &       &  1.7 &   8.8 &   2.6 &   8.8 &   1.9 & 14.2 \\
MC modeling       &  1.3 & 10.5 &  0.2 & 11.7 &  0.6 &  6.4 &  0.5 &  5.7 &       &  0.3 &  14.6 &   4.4 &   8.9 &   1.6 &  8.7 \\
MC statistics     &  0.2 &  2.2 &  2.8 &  9.9 &  0.4 &  5.2 &  0.0 &  1.0 &       &  4.2 &  12.1 &   2.6 &  11.0 &   2.0 &  4.2 \\
Bkg normalization &  1.8 &  5.6 &  3.8 &  7.1 &  2.9 &  6.2 &  2.0 &  6.5 &       &  9.4 &  36.5 &  10.7 &  22.7 &  11.1 & 26.3 \\
\ttbar modeling   &  1.1 & 21.6 &  3.3 & 12.7 &  1.7 & 12.7 &  1.5 & 11.0 &       &  3.6 &  23.3 &  10.7 &  28.9 &   3.2 & 23.6 \\
Jet reco.         &  5.2 & 50.1 &  5.6 & 10.6 &  5.0 & 20.5 &  6.5 & 21.9 &       &  9.0 &  32.9 &  14.5 &  23.5 &  11.3 & 20.9 \\
$b$-tagging       &  \multicolumn{8}{c}{}                                 &       &  4.9 &  33.0 &   5.8 &  11.9 &   6.6 & 10.6 \\
Electron reco.    &  0.4 &  1.9 &  1.3 &  7.1 &  2.4 &  7.0 &  0.7 &  2.8 &       &  3.5 &   8.2 &   0.1 &   9.2 &   1.4 &  3.3 \\
Muon reco.        &  1.1 &  9.7 &  2.2 &  9.3 &  0.2 &  1.9 &  0.5 &  5.1 &       &  0.5 &   2.5 &   0.9 &   6.7 &   0.4 &  2.1 \\
Pile-up           &  0.2 &  9.1 &  0.1 &  2.9 &  0.1 &  1.9 &  0.2 &  1.2 &       &  0.6 &  15.5 &   0.0 &   7.1 &   0.8 &  3.8 \\
Luminosity        &  2.9 &  4.4 &  0.3 &  6.7 &  3.4 &  6.7 &  3.6 &  5.1 &       &  4.6 &   7.4 &   3.3 &   6.9 &   4.4 &  6.2 \\
Unfolding closure &  0.0 & 21.3 &  0.0 &  0.0 &  0.0 & 17.9 &  0.0 &  0.0 &       &  0.0 &   0.0 &   0.0 &   0.0 &   0.0 &  0.0 \\
\midrule
Total syst.       &  7.4 & 62.2 & 11.4 & 26.9 & 10.3 & 33.4 &  9.4 & 27.2 &       & 23.2 &  65.6 &  23.9 &  45.1 &  21.7 & 41.0 \\
\bottomrule
\end{tabular}
\label{tab:systs}
\end{table}

\section{Results}
\label{sec:results}
 
The posterior distributions of nuisance parameters common to the electron and
muon channels were unfolded independently and found to be consistent before
performing the simultaneous unfolding shown here.  Other than in the
inclusive-region \mJ distribution, the FBU procedure did not constrain most
systematic-uncertainty nuisance parameters significantly. The nuisance parameter
for the \sherpa vs.\ \mg modeling uncertainty was the exception: this was
constrained in favor of \sherpa to 20--40\% of the original prior width by all
distributions.
The background normalizations changed by at most a few percent, not significantly
modifying the admixture of signal and background predicted by the MC programs, and
least of all in the 2-tag fits.
 
The full multidimensional posterior-probability distribution is the most
complete form of the measurement, but for histogram presentation each bin's
marginal probability distribution is used to define the central value and error
bar; these correspond to the marginal median and marginal central 68\%
confidence range respectively.
 
\begin{figure}
\centering
\subfigure{\includegraphics[width=0.48\textwidth]{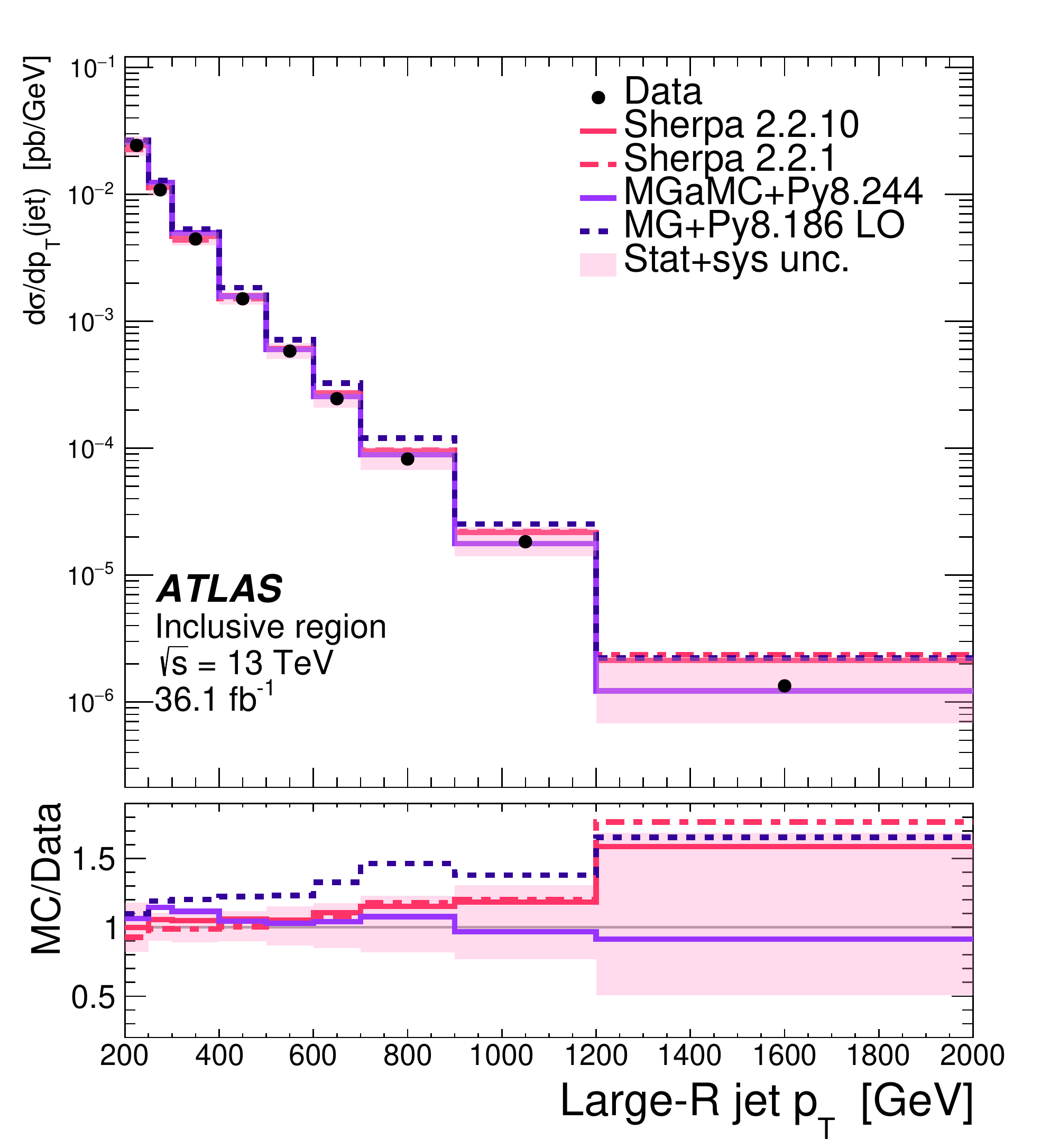}}
\subfigure{\includegraphics[width=0.48\textwidth]{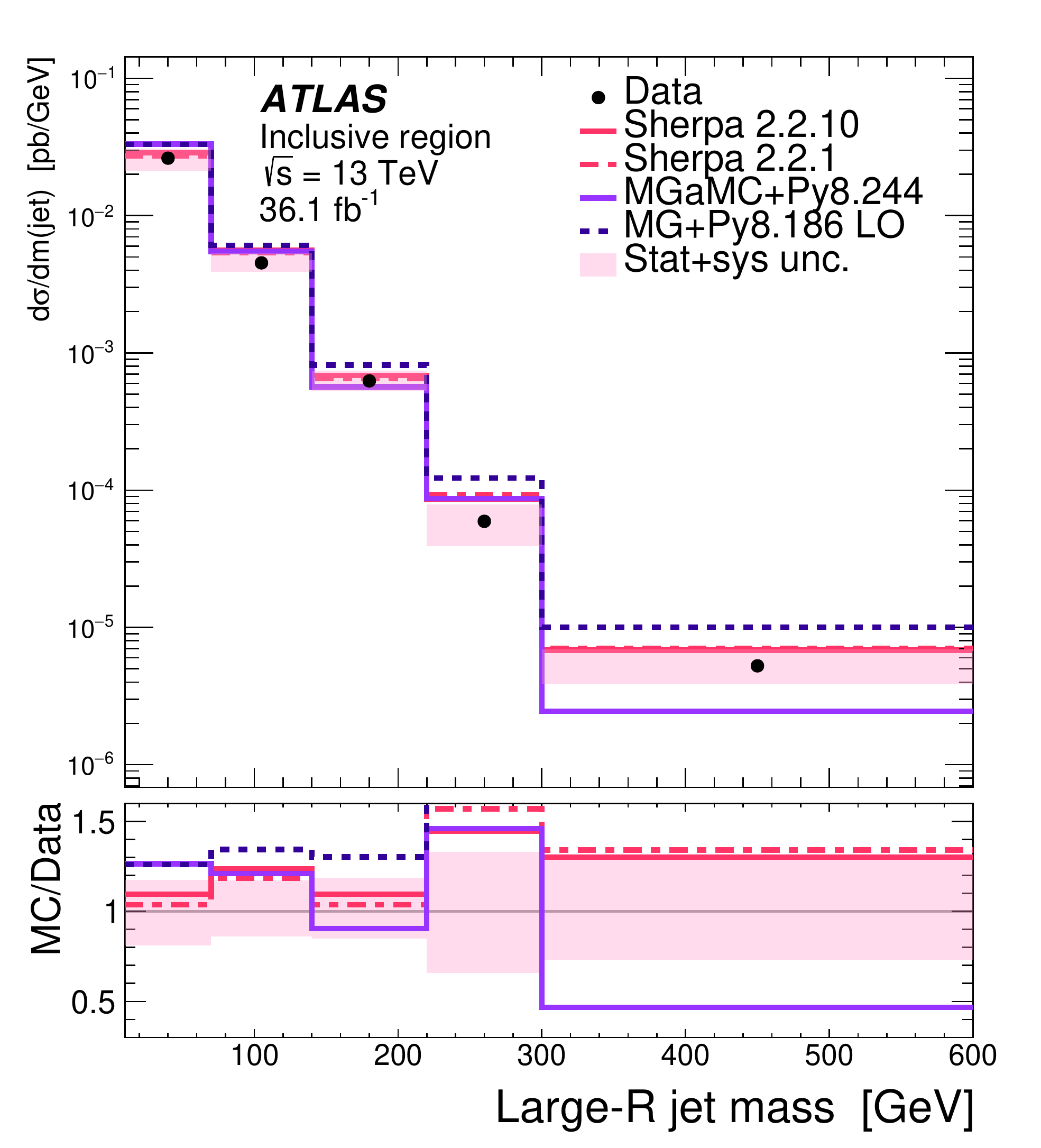}}
\subfigure{\includegraphics[width=0.48\textwidth]{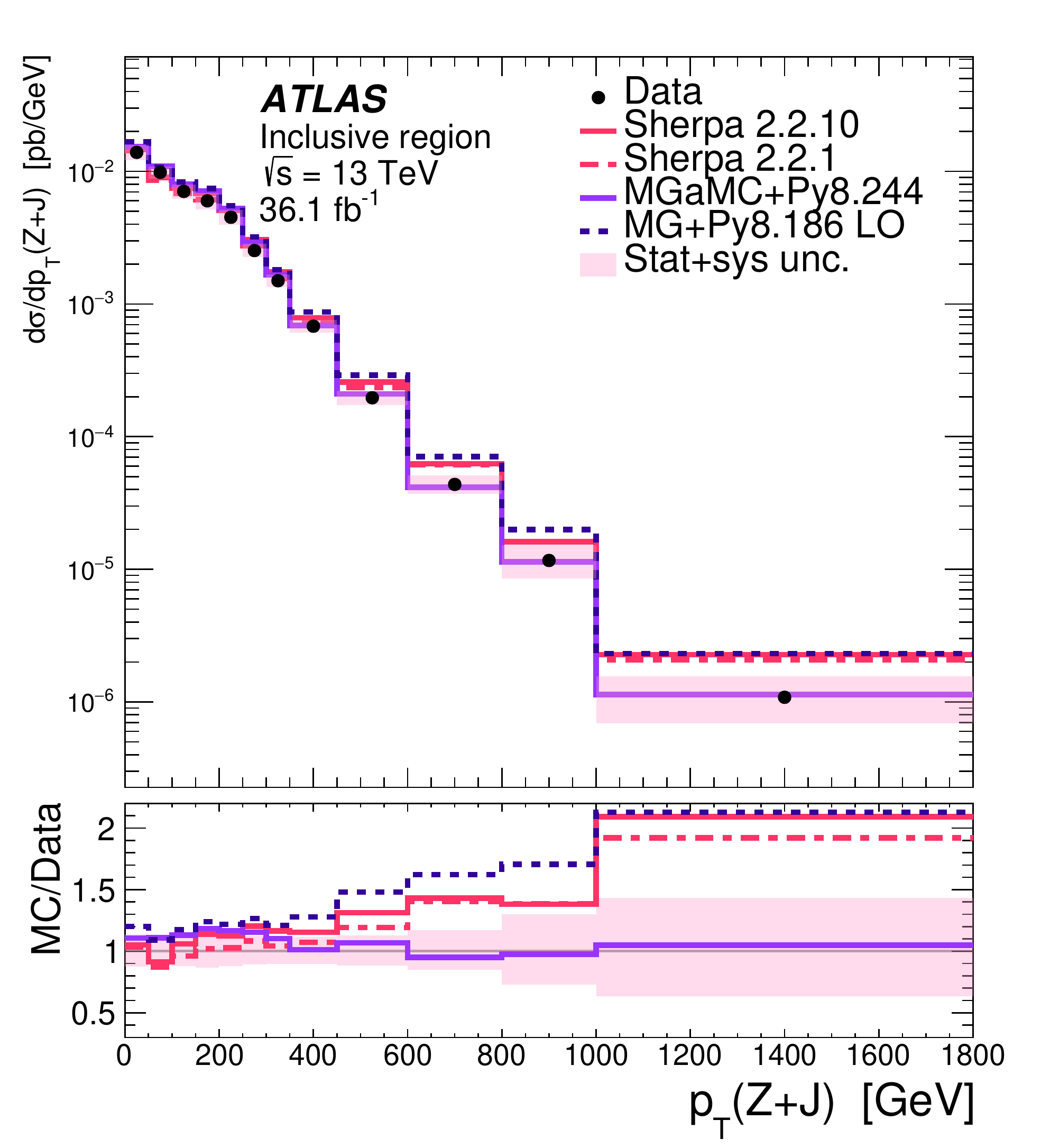}}
\subfigure{\includegraphics[width=0.48\textwidth]{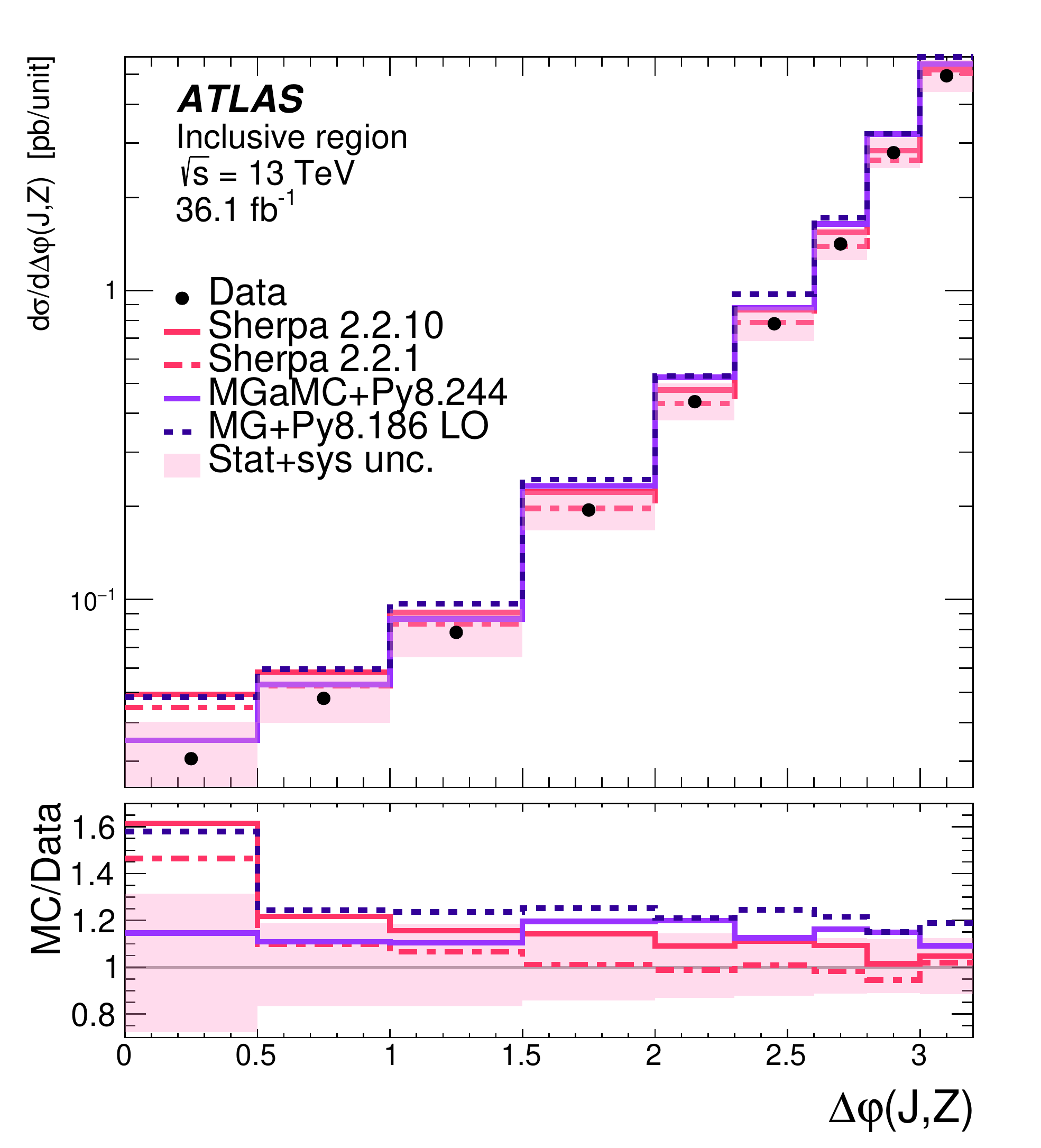}}
\caption{Particle-level differential cross-sections in the inclusive event
selection. The top row shows the \ljet \pt (left) and mass (right),
and the bottom row shows the \pt of the \ZJ system (left) and the azimuthal
separation of the \Z and \ljet (right). The combined statistical and
systematic uncertainty band from the FBU fit is shown.
In the legend,
``MGaMC'' refers to NLO configurations of the \mgamc generator, and ``MG''
to LO \mg, both run in conjunction with \PYTHIAV{8}. All models are using the 5FNS.}
\label{fig:finalplotsincl}
\end{figure}
 
\begin{figure}
\centering
\subfigure{\includegraphics[width=0.48\textwidth]{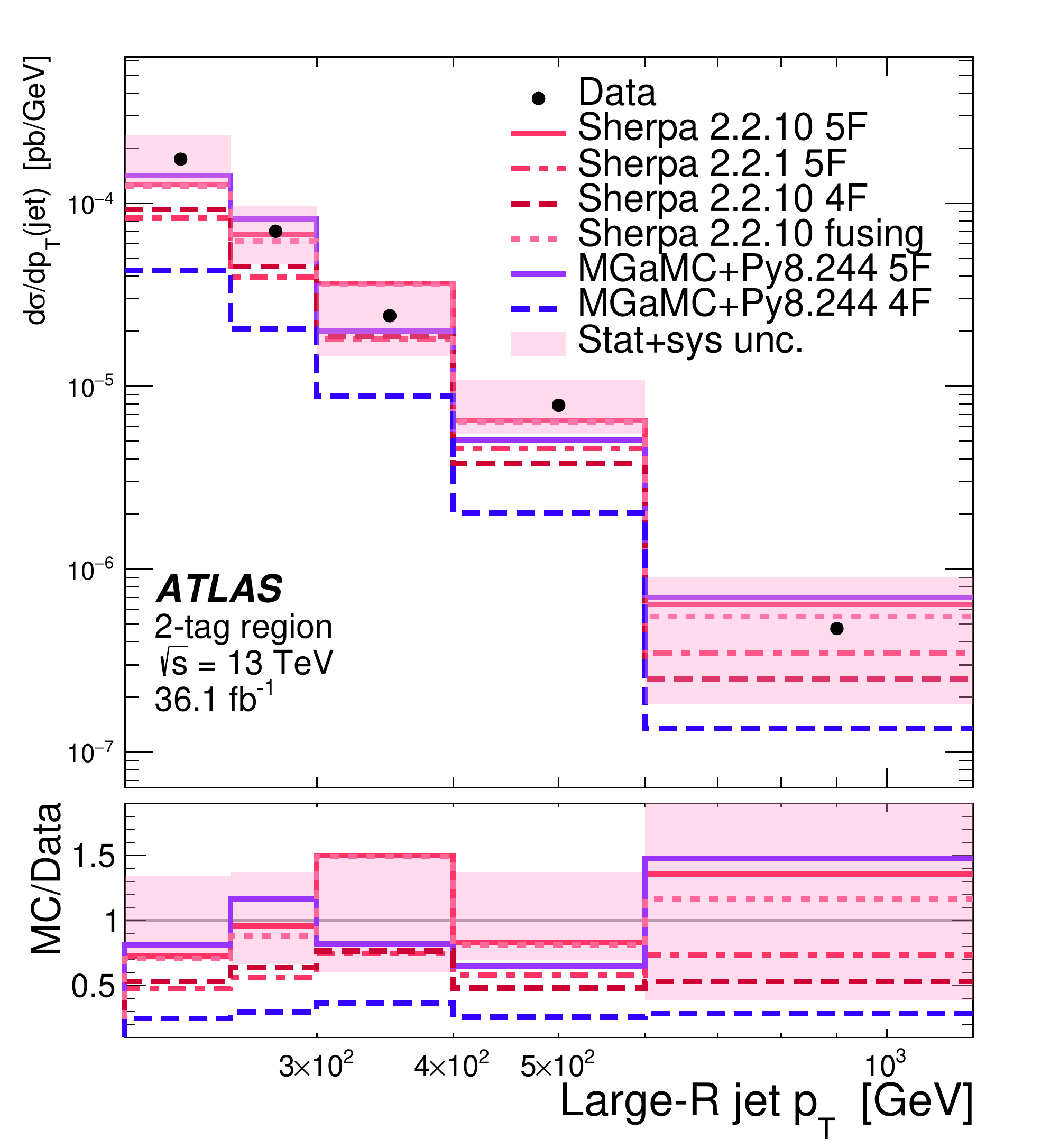}}
\subfigure{\includegraphics[width=0.48\textwidth]{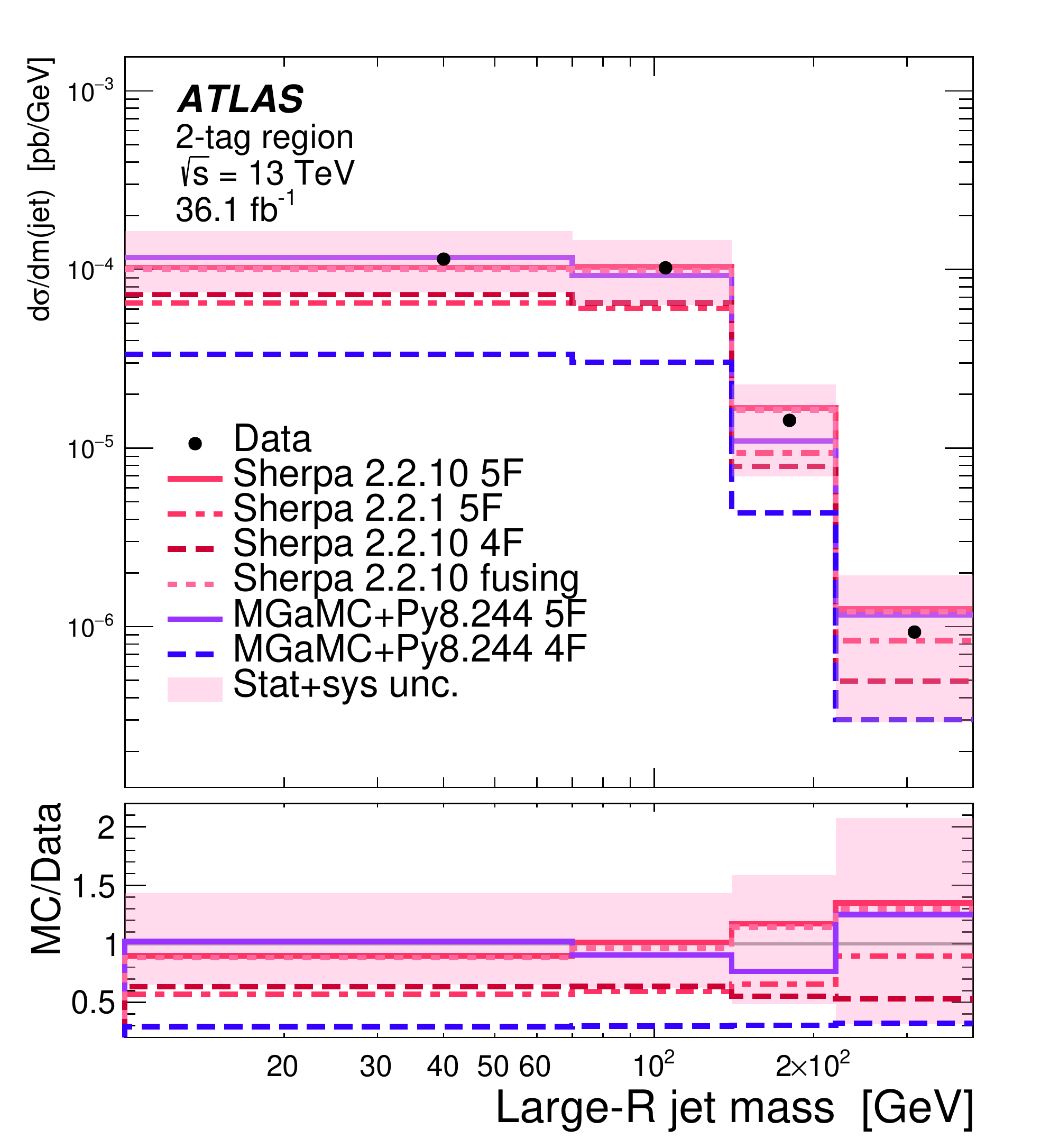}}
\subfigure{\includegraphics[width=0.48\textwidth]{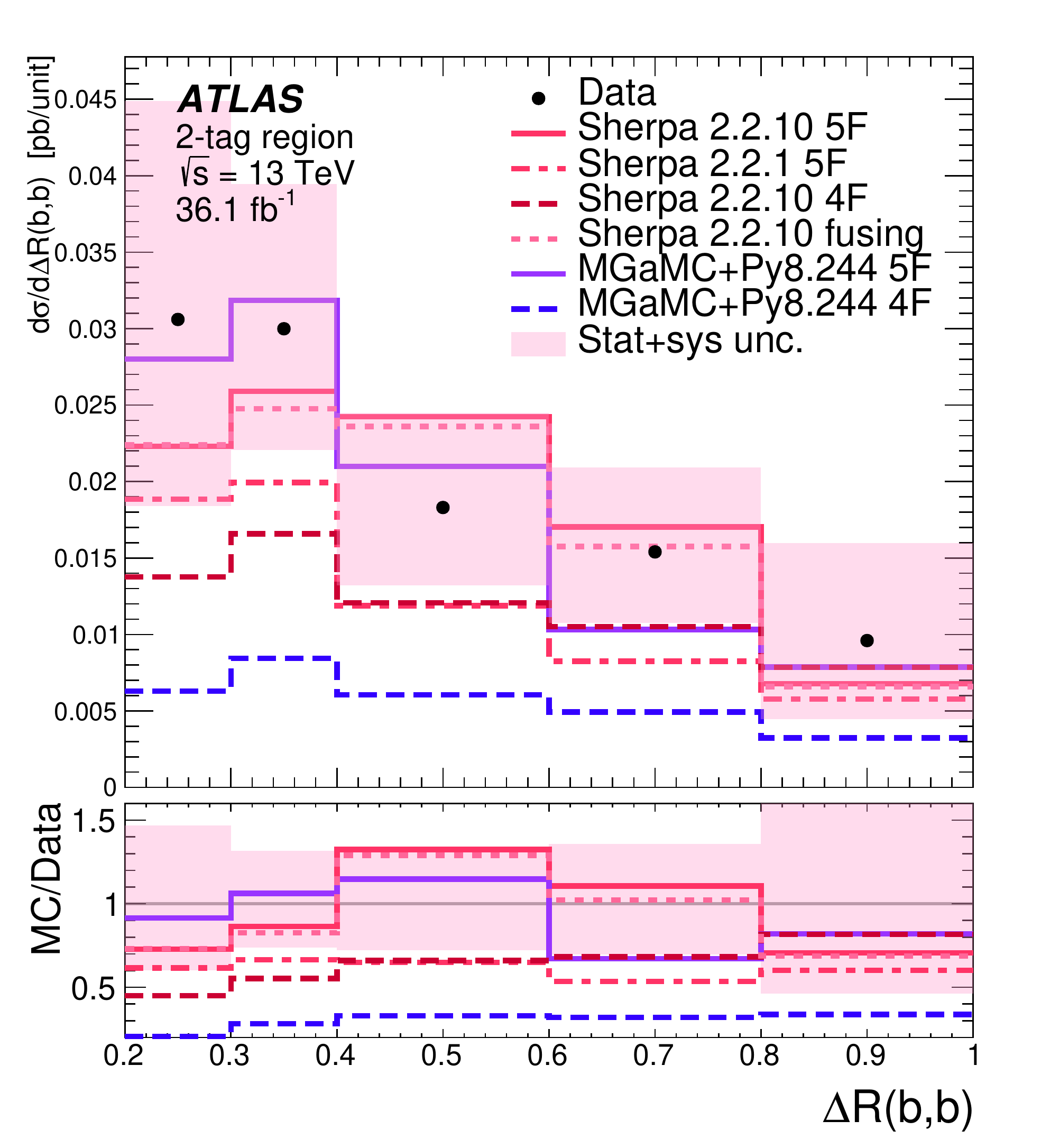}}
\caption{Particle-level differential cross-sections in the 2-tag event
selection. The top row shows the \ljet \pt (left) and mass (right),
and the bottom row shows the angular separation of $b$-tagged
charged-particle subjets. The combined statistical and systematic
uncertainty band from the FBU fit is shown. In the legend, ``MGaMC'' refers
to NLO configurations of the \mgamc generator, run in conjunction with
\PYTHIAV{8}, and ``4/5F'' refer to the flavor-number scheme used.}
\label{fig:finalplots2tag}
\end{figure}

The final unfolded differential cross-sections as functions of the event
kinematics are presented in Figures~\ref{fig:finalplotsincl}
and~\ref{fig:finalplots2tag}, compared with NLO particle-level predictions from
\sherpa~2.2.1 and~2.2.10, and
\mgamc~2.7.3~+~\PYTHIAV{8.244}.  In this section, all predictions are normalized to
their own calculated cross-section to allow an unbiased comparison of both the total
rates and distribution shapes between the different generators.
 
The total fiducial cross-sections are measured by integration of the
angular distributions (chosen because they do not have overflow bins) in both
event-selection regions. The measured values are
$\sigma^\text{incl} = 2.37 \pm 0.28~\text{pb}$ 
for the inclusive selection, and 
$\sigma^\text{2-tag} = 14.6 \pm 4.6~\text{fb}$ 
for the 2-tag selection.

The NLO \sherpa~2.2.10 and NLO \mgamc+~\PYTHIAV{8} generators predict slightly higher central values of the
inclusive cross-section than the measured one, at $2.53 \pm 1.25$~pb
and $2.68 \pm 0.67$~pb respectively, while the LO \mg central configuration
overestimates at $2.84$~pb. The older \sherpa~2.2.1 central prediction of
$2.37$~pb also agrees closely with the measurement. The large uncertainties in
the NLO calculations are dominated by the effects of scale variations;
systematic uncertainties are not well defined for the LO calculation, but the
better nominal performance is evidently provided by the NLO generators.
For the 2-tag cross-section, the NLO 5-flavor \sherpa~2.2.10 and \mgamc central
predictions describe the data well, with $14.9 \pm 4.2$~fb and $14.4 \pm 1.9$~fb
respectively. The \sherpa fusing cross-section, mixing elements of the 4- and
5-flavor calculations, is close to the 5-flavor predictions, with
$14.3 \pm 4.8$~fb. The 4-flavor \sherpa and \mgamc calculations, and the
previous 5-flavor \sherpa~2.2.1 prediction
underestimate the 2-tag cross-section with $9.4 \pm 3.1$~fb, $4.4 \pm 1.1$~fb,
and $9.1$~fb respectively. This result underscores the expectation that the
5-flavor (or fusing) scheme is the more appropriate choice for heavy-quark
production in this analysis phase-space, even for \bbbar-pair production.
 
The $\sigma^\text{2-tag}/\sigma^\text{incl}$ ratio of 2-tag to inclusive events
seen in data is $(0.62 \pm 0.12)\%$, accounting for cancellations of shared systematic
uncertainties between the inclusive and 2-tag cross-section estimates. This
figure is reproduced well by \sherpa~2.2.10's $(0.59 \pm 0.39)\%$ and by \mgamc+~\PYTHIAV{8}
with $(0.54 \pm 0.21)\%$.  The older NLO \sherpa~2.2.1 and leading-order \mg+~\PYTHIAV{8}
estimates undershoot with 0.42\% and 0.38\% respectively. These cross-sections hence
furnish new experimental discriminators between perturbative-QCD models of
high-\pT heavy-flavor production rates, despite the significant measurement
uncertainties.
 
In the inclusive-selection differential distributions of
Figure~\ref{fig:finalplotsincl}, the \mgamc+~\PYTHIAV{8} predictions can be seen to have the
shapes in best agreement with data, not suffering from the excesses of activity
common to the \sherpa models and leading-order \mg+~\PYTHIAV{8} in the more extreme
phase-space regions of high \ptJ and \ptZJ, and small \dphiZJ.  In this
topology, where the \zjets process
becomes more like a dijet system with collinear \Z-boson radiation, both generators display similar
shape deviations with respect to the measurement, with the best agreement at low
\pt, low mass, and low levels of additional event activity (as characterised by
low \ptZJ values and the back-to-back $\dphiZJ \sim \pi$ region). The
excess in inclusive cross-section estimate for nominal \mgamc+\pythia\,8 can be seen to arise from
a relatively small overpopulation with respect to \sherpa in the most populated
bins of \ptJ and \ptZJ, while its shapes typically match data to within 10\%,
whereas the other generators overestimate high-scale activity by 50--100\%.
 
As noted in the review of reconstruction-level plots, the mismodeling of extra
radiation by \sherpa and leading-order \mg+~\PYTHIAV{8} is one of the most significant
discrepancies between simulation and data observed in this analysis. Despite the
large measurement uncertainties, this evidence of larger transverse recoil of
the \ZJ system in simulation than in data, as well as the higher-than-observed
\pt and mass of the \ljet{s}, is an important result for inclusive QCD model
development and tuning in this boosted phase-space.
 
In the 2-tag selection distributions in Figure~\ref{fig:finalplots2tag}, the
larger uncertainties and much lower event counts mean that shape discrepancies
are more difficult to discern: current shape modeling appears to
be performing adequately, with relatively constant MC/data ratios for \ljet mass and
\drbb. In particular, the \ljet mass in this region appears to be consistently
well described by all MC models, with no sign of the excesses and model
disagreements seen in the inclusive-region version of that observable. Further
analysis with the complete Run~2 dataset will be required to discriminate
between the models in this phase-space, beyond the evident favoring of the 5FNS
for total \bbbar production rate.
 
\FloatBarrier

\section{Conclusion}
\label{sec:conclusion}
 
We have presented measurements of cross-sections for the production of a
leptonically decaying \Z boson in association with a large-radius jet in LHC
13~\TeV{} proton--proton collision events from the ATLAS $36~\text{fb}^{-1}$
combined 2015--2016 data-taking run, corrected to a particle-level fiducial
region. The observables presented are differential in kinematic variables of the
\Z boson, the large-$R$ jet, and its associated small-$R$ $b$-tagged charged-particle
jets. They are measured with a flavor-inclusive event selection and also within
a ``2-tag'' event-selection region which adds a double
$b$-hadron labeling requirement on the \ljet.  The integrated cross-sections
within the fiducial volumes of the event-selection regions have also been presented.
 
These cross-section estimates were extracted from data using the Fully Bayesian
Unfolding formalism, effectively performing a posterior-probability fit over a
combination of signal and background cross-section parameters, and various
systematic uncertainties affecting the response of the detector.
These measurements provide an important test of perturbative quantum
chromodynamics, with particular emphasis on the production rates and kinematics
of bottom quarks. These are a significant background to several important
Higgs-boson searches, and are affected by significant theory and modeling
uncertainties. The full data, correlations,
and samples from the posterior-probability function are provided for use
in event-generator tuning and model hypothesis-testing
via public databases.

The differential cross-sections indicate significant mismodeling of QCD
activity in the inclusive event selection by many MC models, with both the NLO
\sherpa and LO \mg+~\PYTHIAV{8} event generators predicting greater \pt and azimuthal
decorrelation in the \ZJ system than seen in the ATLAS data. The \ljet itself is
consequently biased to higher \pt and mass values than in data, although to a
lesser extent than the deviations in the \ZJ-system observables. The NLO \mgamc+~\PYTHIAV{8}
model, by contrast, describes all distribution shapes well, with only a small
overestimate of the inclusive fiducial cross-section. All models somewhat
overestimate this cross-section, with recent \sherpa versions providing the best
description.
 
The 2-tag selection, while its discrimination power is limited by the number of data
events, does not appear to suffer from the same shape-modeling issues, and there is
good shape agreement between the data and all MC models. The strongest feature
observed in this event-selection region is in normalization, with models using
the 4FNS approach significantly underestimating the rate of \bbbar boosted-jet
production.  Five-flavor approaches with modern tools do much better, with both
\sherpa~2.2.10 and \mgamc providing accurate predictions for the 2-tag
cross-section and the ratio of 2-tag to inclusive rates.
This information is important for future use of MC-derived \ljet flavor
composition in, for example, studies of the $\VH(\bbbar)$ process.  As the
result is statistically limited in the 2-tag region, the significant increase in
integrated luminosity from the LHC full Run~2 dataset (and expected from the LHC
Run~3 program) should provide a clearer view of how far the validity of MC
modeling of heavy-flavor extends into this extreme event topology.

\section*{Acknowledgments}

% The next lines are included from the .//acknowledgements/Acknowledgements.tex input file

We thank CERN for the very successful operation of the LHC, as well as the
support staff from our institutions without whom ATLAS could not be
operated efficiently.
 
We acknowledge the support of
ANPCyT, Argentina;
YerPhI, Armenia;
ARC, Australia;
BMWFW and FWF, Austria;
ANAS, Azerbaijan;
CNPq and FAPESP, Brazil;
NSERC, NRC and CFI, Canada;
CERN;
ANID, Chile;
CAS, MOST and NSFC, China;
Minciencias, Colombia;
MEYS CR, Czech Republic;
DNRF and DNSRC, Denmark;
IN2P3-CNRS and CEA-DRF/IRFU, France;
SRNSFG, Georgia;
BMBF, HGF and MPG, Germany;
GSRI, Greece;
RGC and Hong Kong SAR, China;
ISF and Benoziyo Center, Israel;
INFN, Italy;
MEXT and JSPS, Japan;
CNRST, Morocco;
NWO, Netherlands;
RCN, Norway;
MEiN, Poland;
FCT, Portugal;
MNE/IFA, Romania;
MESTD, Serbia;
MSSR, Slovakia;
ARRS and MIZ\v{S}, Slovenia;
DSI/NRF, South Africa;
MICINN, Spain;
SRC and Wallenberg Foundation, Sweden;
SERI, SNSF and Cantons of Bern and Geneva, Switzerland;
MOST, Taipei;
TENMAK, T\"urkiye;
STFC, United Kingdom;
DOE and NSF, United States of America.
In addition, individual groups and members have received support from
BCKDF, CANARIE, CRC and DRAC, Canada;
PRIMUS 21/SCI/017 and UNCE SCI/013, Czech Republic;
COST, ERC, ERDF, Horizon 2020, ICSC-NextGenerationEU and Marie Sk{\l}odowska-Curie Actions, European Union;
Investissements d'Avenir Labex, Investissements d'Avenir Idex and ANR, France;
DFG and AvH Foundation, Germany;
Herakleitos, Thales and Aristeia programmes co-financed by EU-ESF and the Greek NSRF, Greece;
BSF-NSF and MINERVA, Israel;
Norwegian Financial Mechanism 2014-2021, Norway;
NCN and NAWA, Poland;
La Caixa Banking Foundation, CERCA Programme Generalitat de Catalunya and PROMETEO and GenT Programmes Generalitat Valenciana, Spain;
G\"{o}ran Gustafssons Stiftelse, Sweden;
The Royal Society and Leverhulme Trust, United Kingdom.
 
The crucial computing support from all WLCG partners is acknowledged gratefully, in particular from CERN, the ATLAS Tier-1 facilities at TRIUMF/SFU (Canada), NDGF (Denmark, Norway, Sweden), CC-IN2P3 (France), KIT/GridKA (Germany), INFN-CNAF (Italy), NL-T1 (Netherlands), PIC (Spain), RAL (UK) and BNL (USA), the Tier-2 facilities worldwide and large non-WLCG resource providers. Major contributors of computing resources are listed in Ref.~\cite{ATL-SOFT-PUB-2023-001}.

% End of text imported from the .//acknowledgements/Acknowledgements.tex input file

\printbibliography
 
\clearpage
% ATLAS Collaboration author list
% Reference date of STDM-2017-37 is 2020-08-28
% Author list last updated on date 12-JAN-24
% Data extracted on 12-Jan-2024 for paper reference STDM-2017-37
% at 4:17pm
 
\begin{flushleft}
\hypersetup{urlcolor=black}
{\Large The ATLAS Collaboration}

\bigskip

\AtlasOrcid[0000-0002-6665-4934]{G.~Aad}$^\textrm{\scriptsize 103}$,
\AtlasOrcid[0000-0002-5888-2734]{B.~Abbott}$^\textrm{\scriptsize 121}$,
\AtlasOrcid[0000-0002-7248-3203]{D.C.~Abbott}$^\textrm{\scriptsize 104}$,
\AtlasOrcid[0000-0002-2788-3822]{A.~Abed~Abud}$^\textrm{\scriptsize 36}$,
\AtlasOrcid[0000-0002-1002-1652]{K.~Abeling}$^\textrm{\scriptsize 55}$,
\AtlasOrcid[0000-0002-2987-4006]{D.K.~Abhayasinghe}$^\textrm{\scriptsize 95}$,
\AtlasOrcid[0000-0002-8496-9294]{S.H.~Abidi}$^\textrm{\scriptsize 29}$,
\AtlasOrcid[0000-0002-8279-9324]{O.S.~AbouZeid}$^\textrm{\scriptsize 42}$,
\AtlasOrcid{N.L.~Abraham}$^\textrm{\scriptsize 147}$,
\AtlasOrcid[0000-0001-5329-6640]{H.~Abramowicz}$^\textrm{\scriptsize 152}$
% ...
\bigskip
\\

% ...
$^{29}$Physics Department, Brookhaven National Laboratory, Upton NY; United States of America.\\
% ...
$^{36}$CERN, Geneva; Switzerland.\\
% ...
$^{42}$Niels Bohr Institute, University of Copenhagen, Copenhagen; Denmark.\\
% ...
$^{55}$II. Physikalisches Institut, Georg-August-Universit\"{a}t G\"ottingen, G\"ottingen; Germany.\\
% ...
$^{95}$Department of Physics, Royal Holloway University of London, Egham; United Kingdom.\\
% ...
$^{103}$CPPM, Aix-Marseille Universit\'e, CNRS/IN2P3, Marseille; France.\\
$^{104}$Department of Physics, University of Massachusetts, Amherst MA; United States of America.\\
% ...
$^{121}$Homer L. Dodge Department of Physics and Astronomy, University of Oklahoma, Norman OK; United States of America.\\
% ...
$^{147}$Department of Physics and Astronomy, University of Sussex, Brighton; United Kingdom.\\
% ...
$^{152}$Raymond and Beverly Sackler School of Physics and Astronomy, Tel Aviv University, Tel Aviv; Israel.\\
% ...

\end{flushleft}

% Created with Glance <Atlas.Glance@cern.ch>

\end{document}